\documentclass[aps,pre,amsmath,amsfonts,floatfix,superscriptaddress,nofootinbib]{revtex4}

\usepackage{amsmath,amssymb,color,bm}
  \usepackage{amsthm}
  \usepackage{graphicx}

\usepackage{mathrsfs} \usepackage{color} \usepackage{multirow}
\usepackage{bbm}  \usepackage{commath} 
\usepackage{comment}

\let\a=\alpha \let\b=\beta \let\g=\gamma \let\d=\delta
 \let\z=\zeta \let\h=\eta 
\let\l=\lambda \let\m=\mu \let\n=\nu  \let\p=\pi
\let\s=\sigma \let\t=\tau \let\f=\varphi 
   \let\G=\Gamma
\let\D=\Delta \let\Th=\Theta \let\X=\Xi 
   
\let\ee=\varepsilon \let\r=\rho \let\th=\theta \let\io=\infty
\let\om=\omega
\def\ie{{i.e. }}\def\eg{{e.g. }}

\def\EE{{\cal E}}\def\MM{{\cal M}} 
\def\FF{{\cal F}} 
 
\def\LL{{\cal L}}  \def\OO{{\cal O}}
\def\AA{{\cal A}} 
\def\KK{{\cal K}}

  \def\erf{\text{erf}}

\def\redv{\bar v}

\def\wee{\wh\ee}

\def\wfd{{\wh\f}_{\rm d}}

\def\rr{\mathbf{r}}

\def\de{\mathrm d}

\def\max{\mathrm{max}}
\def\min{\mathrm{min}}

\def\to{\rightarrow} \def\la{\left\langle} \def\ra{\right\rangle}

\newcommand{\beq}{\begin{equation}} \newcommand{\eeq}{\end{equation}}
\newcommand{\wh}{\widehat} \newcommand{\wt}{\widetilde}

\newcommand{\afunc}[1]{\operatorname{\mathsf{#1}}}
\def\DE{\afunc{D}}

\begin{document}

\title{Numerical solution of the dynamical mean field theory \\  of infinite-dimensional equilibrium liquids}

\author{Alessandro Manacorda}
\email{alessandro.manacorda@phys.ens.fr}
\affiliation{Laboratoire de Physique de l'Ecole Normale Sup\'erieure, ENS, Universit\'e PSL, CNRS, Sorbonne Universit\'e, Universit\'e de Paris, F-75005 Paris, France}
\author{Gr\'egory Schehr}
\affiliation{Universit\'e Paris-Saclay, CNRS, LPTMS, 91405, Orsay, France}
\author{Francesco Zamponi}
\affiliation{Laboratoire de Physique de l'Ecole Normale Sup\'erieure, ENS, Universit\'e PSL, CNRS, Sorbonne Universit\'e, Universit\'e de Paris, F-75005 Paris, France}

\begin{abstract}
We present a numerical solution of the dynamical mean field theory of infinite-dimensional equilibrium liquids
established in [Phys.~Rev.~Lett. {\bf 116}, 015902 (2016)].
For soft sphere interactions, we obtain the numerical solution by an iterative algorithm and a straightforward discretization of time. We also discuss the case of hard spheres, for which we first derive analytically the dynamical mean field theory as a non-trivial limit of that of soft spheres. We present numerical results for the memory function and the mean square displacement. Our results reproduce and extend kinetic theory in the dilute or short-time limit, while they also describe dynamical arrest towards the glass phase in the dense strongly-interacting regime.
\end{abstract}

\maketitle

\tableofcontents

\section{Introduction}
\label{sec:intro}

Solving the dynamics of dense equilibrium liquids is a notoriously difficult problem~\cite{Hansen}. In the 
low-density limit, or at short times, the solution is obtained via kinetic theory, which allows one to obtain microscopic expressions
for the correlation functions and transport coefficients, while at any density, hydrodynamics provides a description of the 
dynamics at large length and time scales taking the transport coefficients as 
input.
However, at high densities or low temperatures, kinetic theory breaks down, 
and the hydrodynamic regime is pushed to scales that are much larger than 
any experimentally relevant scale. In this strongly interacting ``supercooled 
liquid'' regime, dynamics is slow, viscosity is high, and both correlation 
functions and transport coefficients display non-trivial behavior that is not 
captured by kinetic theory. The only microscopic description of this regime is 
obtained by the Mode-Coupling Theory (MCT)~\cite{BGS84,Go09}, which is a 
set of closed equations derived from a series of poorly controlled 
approximations of the true dynamics. Despite its non-systematic 
nature~\cite{DM86,ABL06,KiKa07,JvW11}, MCT accurately describes the initial 
slowing down of liquid dynamics upon supercooling, including the wavevector 
dependence of correlation functions~\cite{Go99}.

Interestingly, in the formal limit in which the spatial dimension $d$ goes to infinity, liquid thermodynamics reduces to the calculation of the second virial coefficient~\cite{FRW85,FP87,WRF87,FP99}.
Based on this observation, a first attempt to solve exactly liquid dynamics for $d\to\io$ was presented in~\cite{KW87} (see also~\cite{EF88}), but the full dynamical mean field theory (DMFT) that describes exactly the equilibrium dynamics in $d\to\io$ was only derived recently, via a second-order virial expansion on trajectories~\cite{MKZ15} or via a dynamic cavity method~\cite{Sz17,ABUZ18,AMZ18}.
The DMFT provides a set of closed one-dimensional integro-differential equations,
which exactly describe the many-body liquid dynamics in the thermodynamic limit, and are similar in structure to those obtained for quantum systems in the same limit~\cite{GKKR96}.
We refer the reader to~\cite{CKPUZ17,AMZ18,AMZ19,PUZ20} for a detailed review of the solution of liquid dynamics in infinite dimensions, including its extension to the out-of-equilibrium setting (see also~\cite{PLW19} for a related approach).

Unfortunately, the analytical solution of the DMFT equations is out of reach. 
In this work, we present their numerical solution, obtained through an iterative method and a straightforward discretization of time. This strategy, however, only works for differentiable interaction potentials. We thus discuss how to derive the DMFT of hard spheres via a non-trivial limit of a soft sphere interaction. We present numerical results for soft and hard spheres, supported by analytical 
computations at low densities, in the short time limit, and at long times in the glass phase. 

The paper is organized as follows.
In section~\ref{sec:DMFT} we review the basic equations of the DMFT of liquids. In section~\ref{sec:HS}, we discuss the hard sphere limit of DMFT. 
In section~\ref{sec:num}, we discuss the discretization and convergence algorithms used in this work. In section~\ref{sec:results}, we present the numerical solution for soft and hard spheres. Finally, in section~\ref{sec:conclusions} we draw our conclusions and present some perspectives for future work. A few technical discussions are presented in Appendix.
Note that the theoretical analysis developed in sections~\ref{sec:DMFT} and~\ref{sec:HS}
and the numerical methods and results reported in sections~\ref{sec:num} 
and~\ref{sec:results} can also be read independently; we present them jointly for the sake of completeness.

\section{Dynamical mean field theory}
\label{sec:DMFT}

\subsection{General formulation}
\label{sec:DMFE}

In the following, we consider a liquid of interacting particles in $d$ dimensions, 
with pair interaction potential $v(r)$ having a typical interaction scale $\ell$,
at temperature $T$ and number density $\r$ in the thermodynamic limit.
We denote by $\wh\f = 2^d \f / d = \r V_d \ell^d/d$ the scaled packing fraction, 
where $V_d = \pi^{d/2}/\G(1+d/2)$ is the volume of a unit sphere in $d$ dimensions. 
In the infinite dimensional limit, it is convenient to describe the dynamics in terms 
of the inter-particle gap $h = d(r/\ell-1)$, defining as well the rescaled potential 
$\redv(h) = v[\ell (1+h/d)]$, which is assumed to have a finite limit when $d\to\io$~\cite{PUZ20}.

In the most general case,
we consider the Langevin dynamics of the system, being $\wh m = (\ell^2/2d^2)m$ 
and $\wh\z = (\ell^2/2d^2)\z$ respectively the scaled mass and friction coefficient, which are kept finite
in the limit $d\to\io$ in order to obtain a non-trivial dynamics~\cite{MKZ15,AMZ18,PUZ20}.
Denoting a time derivative by a dot,
$\dot h(t) = \de h/\de t$,
DMFT leads
to the following set of self-consistent equations, which describe exactly the equilibrium dynamics when $d\to\io$~\cite{MKZ15,Sz17,AMZ18,PUZ20}:
\begin{equation}\label{eq:DYN}
\begin{split} 
\wh m \ddot h(t) + \wh\z \dot h(t) 
&= T- \redv'(h(t)) - \b \int_{0}^t \de u \,   \MM(t-u) \dot h(u) + \X(t)  \ ,\\
h(t=0) &= h_0 \ , \\
\dot h(t=0) &= {\dot h_0} \ , \\
\langle \X(t) \X(u) \rangle  &= 2 \wh\z T \d(t-u) +   \MM(t-u) \ ,
\end{split}\end{equation}
where $\Xi(t)$ is a colored Gaussian noise with zero mean and a memory kernel $\MM(t)$ self-consistently determined by
\beq\label{eq:Mdef}
\MM(t)   
=\frac{\wh\f}{2} \int \de {\dot h_0} \, \sqrt{\frac{\b \wh m}{2 \pi}} e^{-\b \wh m {\dot h_0}^2 / 2} \int \de h_0 \, e^{h_0 -\b \redv(h_0)} \redv'(h_0) \langle \redv'(h(t))  
\rangle_{h_0, {\dot h_0}}
\ ,
\eeq
where $\langle \rangle_{h_0,{\dot h_0}}$ is the average over the noise in the dynamical process starting 
in $h_0$ with velocity ${\dot h_0}$ (see Appendix~\ref{app:Maxwell} for a discussion of the initial
distribution of $\dot h_0$). The integrals over $h_0$ and $\dot h_0$ are performed over the real axis, 
as it will be implicitly understood for all the following integrals, whenever the integration bounds are not explicitly specified.

From the knowledge of the memory kernel $\MM(t)$ one can derive all the dynamical observables, such as the scaled mean square displacement,
$\frac{1}N \sum_i \la |\mathbf{r}_i(t)-\mathbf{r}_i(0)|^2 \ra = \ell^2 \D(t)/d$, which is the solution 
of the equation
\beq
\label{eq:MSD}
\wh m \ddot \Delta(t) + \wh\z \dot\Delta(t)  = T -\b\int_{0}^t \de u \; \MM(t-u) \dot \Delta(u) \ ,
\eeq
the associated scaled diffusion constant
\beq\label{eq:D}
\wh D = \frac{2d^2}{\ell^2} D = \lim_{t\to\io} \frac{\Delta(t)}t =  \frac{T}{\wh\z + \b \int_0^\io \de t \, \MM(t)} \ ,
\eeq
and the scaled shear viscosity (see Appendix~\ref{app:etas})
\beq\label{eq:etas}
\frac{\wh\h_s}{\wh\f} = \frac{\h_s}{d \, \r}
= \frac{\b \wh m^2}d \int_0^\io \de t \,
\ddot \D(t)^2 + \b \int_0^\io \de t\, \MM(t) \ .
\eeq
We note that the kinetic contribution, i.e. the first term in Eq.~\eqref{eq:etas}, is subleading when $d\to\io$ at constant density. However, it also diverges in the dilute limit $\r\to0$, in which the motion becomes ballistic, so it will be useful to keep this term to compare the DMFT results with finite-dimensional simulation data and with kinetic theory (see section~\ref{ssec:results-HSN}). 
We emphasize that the limit $d \to \io$ with finite $\wh\f$ implies that $\rho 
\to \io$, because the volume of the $d$-dimensional hypersphere vanishes 
faster than exponentially upon increasing $d$. For this reason, the two limits ($\r\to0$ followed by $d\to\io$ and $d\to\io$ followed by $\wh\f\to0$) 
do not  commute, and we need to keep $d$ finite when sending $\r$ (or $\wh\f$) 
to zero to observe the low-density divergence of the scaled viscosity. 
We also stress that the divergence occurs only for $\wh\h_s/\wh\f \propto \h_s/\r$, 
while the low density limit of the shear viscosity $\h_s$ (or $\wh\h_s$) is finite, as predicted \eg by 
the Boltzmann equation.

The DMFT Eqs.~\eqref{eq:DYN} and \eqref{eq:Mdef} involve a 
potential term $\redv'(h)$, a non-Markovian memory contribution from the 
integral term, and a Gaussian colored noise $\Xi(t)$. The goal of the present 
study is to determine  $\MM(t)$ as a self-consistent solution of these equations. 
The knowledge of $\MM(t)$ is the 
fundamental step to compute the dynamical observables of the system as a function of the control parameters $\wh\f, T$. 

The physics of the system also strongly depends on the potential $\redv(h)$. 
In the following, we will restrict ourselves to three kind of potentials: 
1) a linear soft sphere (SLS) potential, $\redv_{\rm SLS}(h) = \ee h \, \theta(-h)$; 
2) a quadratic soft sphere (SQS) potential, $\redv_{\rm SQS}(h) = \ee h^2 \th (-h) /2$; 
3) a hard sphere potential (HS),
$\exp[-\redv_{\rm HS}(h)] = \th(h)$, 
being $\th(x)$ the Heaviside step function.
All those potentials are short-ranged, with vanishing interaction for $h>0$ (no overlap 
between particles), they are purely repulsive and both SLS and SQS tend to HS in the 
limit $\ee \to \io$. The energy scale $\ee$ gives the strength of the interactions, and 
its ratio with the temperature $T$ will be one of the dimensionless control parameters of our model.

It is clear that the DMFT equations, as written in Eqs.~\eqref{eq:DYN} 
and~\eqref{eq:Mdef}, cannot be straightforwardly applied to the HS case, in which the potential 
is not differentiable and the $\redv'(h)$ terms are ill-defined. In section~\ref{sec:HS}, we will show 
how to regularize these equations to have a well-defined memory function $\MM(t)$.
The introduction of two kinds of soft sphere potentials (SLS and SQS) is motivated by the need of simple 
analytical calculations and unambiguous numerical solutions. Indeed, in section~\ref{sec:HS} 
we will show how the short-time or low-density limit of $\MM(t)$ for hard spheres can be 
approached analytically, using the SLS potential for the sake of simplicity. Viceversa, the
numerical solutions shown in section~\ref{ssec:results-SS} have been found with a SQS potential. 
The latter has a regular derivative in $h=0$ - at variance with the SLS potential - 
and, therefore, the force changes continuously when particles get in contact and the numerical 
results are much clearer. Of course, the HS limit should not depend on the particular soft spheres 
potential chosen.

\subsection{Dimensionless equations}
\label{ssec:adim}

We now briefly discuss what are the dimensionless parameters that control the dynamical behavior, in the Brownian and Newtonian case, obtained as particular limits of the general Langevin dynamics.
The discussion of the general (mixed) case is a straightforward extension of the ones below and is not reported for conciseness.

\subsubsection{The Brownian case}
\label{sssec:adim-B}

In the Brownian (overdamped) case, one takes $\wh m = 0$; the evolution 
Eq.~\eqref{eq:DYN} becomes then a first-order differential equation, so 
the initial condition on the velocity ${\dot h_0}$ disappears 
from the dynamics and from $\MM(t)$ in Eq.~\eqref{eq:Mdef}.
If $\wh m = 0$, the characteristic time of Eq.~\eqref{eq:DYN} reads $\t_B = 
\wh\z / T$, which will be set to 1 in this case with the rescaling $t/\t_B \rightarrow t$. 
The dimensionless potential, memory kernel and noise are also rescaled as $ \beta \, \redv(h) 
\rightarrow \redv(h)$, $\beta^2 \MM(t \, \t_B) \rightarrow \MM(t)$ and $\beta \, \X(t \, \t_B ) 
\rightarrow \Xi(t)$.
Substituting the rescaled variables into the dynamical evolution 
in Eq.~\eqref{eq:DYN} multiplied by $\beta$ one gets
\begin{equation}\label{eq:DYNadim-B}
\begin{split}
\dot h(t) 
&= 1 - \redv'(h(t)) - \int_{0}^t \de u \,   \MM(t-u) \dot h(u) + \X(t)  \ ,\\
h(t=0) &= h_0 \ , \\
\langle \X(t) \X(u) \rangle  &= 2 \d(t-u) +  \MM(t-u) \ ,
\end{split}\end{equation}
with the dimensionless self-consistent equation:
\beq\label{eq:Mdefadim-B}
\MM(t)   
=\frac{\wh\f}{2}  \int \de h_0 \, e^{h_0 -\redv(h_0)} \redv'(h_0)
\langle \redv'(h(t))  \rangle_{h_0} \ .
\eeq
It is evident that the dynamics is then governed by two dimensionless parameters only: 
the rescaled packing fraction $\wh \f$ and the rescaled interaction strength or 
inverse temperature $\wh \ee = \b \ee$. Note that these rescalings are 
equivalent to setting $\ell=1$, $\t_B=1$ and $\ee=1$ as units of length, time and energy, respectively.

\subsubsection{The Newtonian case}
\label{sssec:adim-N}

Setting $\wh m > 0$ and $\wh\z = 0$ in the 
dynamical Eqs.~\eqref{eq:DYN} gives Newtonian dynamics. 
The dimensionless equations can be obtained 
as in the Brownian case, but the characteristic time now reads $\t_N = 
\sqrt{\wh m / T}$ while the potential, memory and noise scale as in the Brownian 
case. For the initial velocity, one can define $\dot h_0 = g_0 / \t_N$, being $g_0$ 
a Gaussian variable of zero average and unit variance. 
Again, one substitutes the rescaled variables into the Newtonian dynamical Eq.~\eqref{eq:DYN} 
and gets
\begin{equation}\label{eq:DYNadim-N}
\begin{split}
\ddot h(t) 
&= 1 - \redv'(h(t)) - \int_{0}^t \de u \,  \MM(t-u) \dot h(u) + \X(t)  \ ,\\
h(t=0) &= h_0 \ , \\
\dot h (t=0)&= g_0 \ , \\
\langle \X(t) \X(u) \rangle  &= \MM(t-u) \ ,
\end{split}\end{equation}
with the dimensionless self-consistent equation
\beq\label{eq:Mdefadim-N}
\MM(t)   
=\frac{\wh\f}{2}  \int \frac{\de g_0}{\sqrt{2\pi}} e^{-g^2_0/2}   \int \de h_0 \, e^{h_0 -\redv(h_0)} \redv'(h_0)
\langle \redv'(h(t))  \rangle_{h_0, g_0} \ .
\eeq
We emphasize that the Newtonian and Brownian equations have a similar 
structure, but the time scales are totally unrelated, as they depend on different 
physical coefficients. As in the Brownian case, the dimensionless parameters 
characterizing the system are $\wh \f$ and $\wh \ee$. The transformation to dimensionless
equations is now equivalent to setting $\ell=1$, $\t_N=1$ and $\ee=1$.

\subsection{Iterative solution}
\label{ssec:iterative}

As in the DMFT of quantum systems~\cite{GKKR96},
finding $\MM(t)$ for all times $t$ by means of an analytical solution is beyond current 
possibilities. A numerical solution can be found, however, by means of 
the following iterative procedure: 1)~define an initial memory kernel $\MM 
(t) = 0$ (or any other more convenient initial condition); 2)~simulate the stochastic trajectories $h(t)$ from Eq.~\eqref{eq:DYN} 
with $\MM(t)$; 3)~compute the new memory kernel $\MM (t)$ through 
Eq.~\eqref{eq:Mdef}; 4)~repeat until convergence is achieved.
If the potential is differentiable, this procedure can be implemented by a straightforward discretization of time.

The iterative procedure introduced above is, however, not guaranteed to be convergent.
Still, we can observe
that the self-consistent Eq.~\eqref{eq:Mdef} 
takes the form $\MM = \wh\f\FF_{\rm DMFT}[\MM]$, with
an explicit factor of density in front of the implicit functional $\FF_{\rm DMFT}[\MM]$.
Hence, writing the memory kernel in a low density expansion as $\MM = \wh\f \MM^{(1)} 
+ \wh\f^2 \MM^{(2)} + \ldots$, the first-order term $\MM^{(1)}=\wh\f\FF_{\rm DMFT}[0]$ is precisely obtained 
by setting $\MM=0$ in the dynamical process, 
$\MM^{(2)}=\wh\f\FF_{\rm DMFT}[\MM^{(1)}]$ 
is obtained by 
setting $\MM=\MM^{(1)}$, and so on, recursively.
So, the solution of the iterative procedure after $n$ iterations gives the low-density expansion of $\MM(t)$ at order $n$, and the convergence of the iterative procedure is equivalent to the convergence of the low-density expansion.

\subsection{Dynamical glass transition}
\label{ssec:plateau}

The DMFT equations can exhibit a dynamically arrested phase, in which the memory function does not decay to zero but to a finite plateau, and the mean square displacement reaches a plateau without showing any diffusive regime. This ``dynamical glass transition'' belongs to the same universality class as that of MCT.
We briefly recall here how a closed equation for the plateau of $\MM(t)$ can be derived, see~\cite{MKZ15,AMZ18,PUZ20} for details.

Let us assume that the memory can be decomposed into a constant component $\MM_\io$ (the plateau) and a 
``fast'' component $\MM_f(t)$ that decays to zero over a finite time scale, \ie
\beq
\MM(t) = \MM_\io + \MM_f(t) \ , \qquad
\lim_{t\to\io}\MM_f(t)= 0 \ .
\eeq
Correspondingly, we can decompose the noise as $\Xi(t) = \Xi_f(t) + \Xi_\io$, where the two components are independent and have
variance 
$\la \Xi_f(t)\Xi_f(u) \ra = 2 \wh\z T\d(t-u) + \MM_f(t-u)$ and
$\la \Xi_\io^2 \ra = \MM_\io$, respectively.
Plugging this in Eq.~\eqref{eq:DYN} and moving the fast components of the retarded friction and noise to the left hand side, we obtain the general dynamical equation
\beq
\LL[h] \equiv \wh m \, \ddot h(t) + \wh\z \, \dot h(t) + \b \int_0^t \de s \, \MM_f(t-s) \, \dot h(s) - \X_f(t) = T - \redv'(h) - \b \MM_\io (h-h_0) + \Xi_\io \equiv - w'(h) \ ,
\eeq
where $\LL[h]$ is a dynamical operator that encodes all the time derivatives together with the fast components of the noise and friction. Because the noise and friction are in equilibrium at temperature $T$ (\ie they satisfy the fluctuation-dissipation theorem~\cite{Cu02}), at long times the system equilibrates in the potential $w(h)$, thus reaching the conditional distribution
\beq\label{eq:phh0}
p(h|h_0, \Xi_\io) = 
\frac{e^{-\b w(h)}}{\int \de z e^{-\b w(z)}} \ ,
\qquad
w(h) = \redv(h) -T h  +\frac{\b \MM_\io}2 (h-h_0)^2 - \Xi_\io h \ .
\eeq
The presence of a finite plateau adds a harmonic trap to the potential, which confines the particle in a finite region and prevents diffusion.
Hence,
\beq\label{eq:Mplateau}
\lim_{t\to\io}\la \redv'(h(t) \ra_{h_0, \dot h_0} =\int \de \Xi_\io p(\Xi_\io)
\int \de h\, p(h|h_0,\Xi_\io) \redv'(h) \ ,
\eeq
where $p(\Xi_\io)$ is a Gaussian with zero mean and variance $\MM_\io$. 
Taking the limit $t\to\io$ of Eq.~\eqref{eq:Mdef} 
gives a closed equation for $\MM_\io$ in the form
\beq\label{eq:Mio}
\MM_\io
=\frac{\wh\f}{2} \int \de h_0 \, e^{h_0 -\b \redv(h_0)} \redv'(h_0)
\int \de \Xi_\io p(\Xi_\io)\int \de h\, p(h|h_0,\Xi_\io) \redv'(h)
\ .
\eeq
Note that one of the three integrals can be eliminated by some simple manipulation~\cite{Sz17,PUZ20}. At low density or high temperature, Eq.~\eqref{eq:Mio} admits $\MM_\io=0$ as unique solution; at high density or low temperature, instead, a finite solution $\MM_\io>0$ appears, usually in a discontinuous way. The line in the temperature-density plane separating the two situations is called the ``dynamical glass transition line'', $\wh\f_{\rm d}(T)$.
Note that the long time distribution in Eq.~\eqref{eq:phh0} does not depend on the details of the short-time dynamics, encoded in $\LL[h]$, provided the equilibrium conditions are satisfied. As a result, the plateau equation is the same for Brownian and Newtonian dynamics.

\section{Hard sphere limit}
\label{sec:HS}

In this section, we derive the DMFT for hard spheres, by taking the $\wh\ee\to\io$ limit of the DMFT equations for the SLS potential. We discuss separately the Brownian and Newtonian cases because the physical and mathematical properties of the equations are very different in the two cases.

\subsection{Brownian dynamics}
\label{ssec:HSB}

Recall that we work here in dimensionless units as discussed in section~\ref{sssec:adim-B}.
For a Brownian SLS system, the DMFT Eqs.~(\ref{eq:DYNadim-B}-\ref{eq:Mdefadim-B})
become
\beq\label{eq:DYN-SS1-B}
\begin{split}
	\dot h(t)
	&= 1 + \wh\ee \, \th(-h(t)) - \int_{0}^t \de u \,   \MM(t-u) \dot h(u) + \X(t)  \ ,\\
	h(t=0) &= h_0 \ , \\
	\langle \X(t) \X(u) \rangle  &= 2 \, \delta(t-u) + \MM(t-u) \ ,
\end{split}
\eeq
with
\beq\label{eq:M-SS1-B}
\MM(t)   
=\frac{\wh\f}{2}   \int_{-\io}^0 \de h_0 \, e^{h_0 + \wh\ee h_0} \wh\ee^2 G(h_0,t)
\ ,
\qquad
G(h_0,t) = \langle \th(-h(t))  \rangle_{h_0} \ ,
\eeq
where $G(h_0,t)$ is the probability for the dynamical process starting at $h_0$ to end at any $h<0$ 
at time $t$. Note that the factor $\redv'(h_0)$ enforces the condition $h_0<0$ in the memory function.

To gain some insight on the $\wh\ee\to \io$ limit,
we will begin by considering the first iteration of the algorithm described in section~\ref{ssec:iterative}, which corresponds to 
setting $\MM(t) = \MM^{(0)}(t) \equiv 0$ in Eq.~\eqref{eq:DYN-SS1-B}.
We then have
\begin{equation}\label{eq:Brow_drift}
\dot h(t) 
= 1 + \wh\ee \th(-h(t)) + \X(t)  \ , \qquad   \langle \X(t) \X(u) \rangle  = 2 \, \d(t-u) \ ,
\end{equation}
and we want to compute $G(h_0,t)$, which follows the backward Kolmogorov evolution equation (see \eg\cite{BMS13})
\beq\label{eq:Gev}\begin{split}
	&\dot G(h_0,t) = G''(h_0,t) + (1+\wh\ee) G'(h_0,t) \ , \qquad h_0 < 0 \ , \\
	&\dot G(h_0,t) = G''(h_0,t) + G'(h_0,t) \ ,\qquad\qquad\quad h_0 > 0 \ ,
\end{split}\eeq
where primes denote derivatives with respect to $h_0$, and
with boundary conditions
\beq\label{eq:Aboundary}
G(h_0,t=0) = \th(-h_0) \ , \qquad G(h_0\to-\io,t) = 1 \ , \qquad G(h_0\to \io,t) = 0 \ .
\eeq
The evolution Eq.~\eqref{eq:Gev} 
can be solved in Laplace space 
for $h_0<0$ and $h_0>0$ separately.
Imposing the continuity of $G(h_0,t)$ and $G'(h_0,t)$ at $h_0=0$ (see 
Appendix~\ref{app:BrownianHS}), we then obtain
\beq\label{eq:GLap}
\begin{split}
	\wt G(h_0,s) &= \begin{cases}
		\frac1s + c_-(s) e^{\l_-(s) h_0} \ , & h_0 < 0 \ ,     \\
		c_+(s) e^{\l_+(s) h_0}  \ , & h_0 > 0 \ , \\
	\end{cases} \\ 
	\l_-(s) &= \frac1{2} \left( -1 -\wh\ee+\sqrt{(1+\wh\ee)^2+4 s} \right) \ , \qquad  c_-(s) = \frac{\l_+(s)}{s [\l_-(s) - \l_+(s)]}\ , \\
	\l_+(s) &= \frac1{2} \left( -1-\sqrt{1+4 s} \right)  \ , \qquad \qquad \qquad\quad c_+(s) = \frac{\l_-(s)}{s [\l_-(s) - \l_+(s)]} \ . 
\end{split}\eeq
The Laplace transform of $\MM^{(1)}(t)$, corresponding to the first iteration (or first-order in density), is therefore
\beq\label{eq:M1SSlB}
\wt\MM^{(1)}(s)  =\frac{\wh\f}{2}  \int_{-\io}^0 \de h_0 \, e^{h_0 + \wh \ee y_0} \wh \ee^2
 \wt G(h_0,s) = \frac{\wh\f}{2} \wh \ee^2  \left[\frac1{s(1 + \wh\ee)} +\frac{ c_-(s)}{1 + \wh\ee + \l_-(s)} \right] .
\eeq
Note that from $\wt\MM^{(1)}(s)$ we immediately get the lowest order density correction to the diffusion constant via Eq.~\eqref{eq:D}:
\beq
\wt\MM^{(1)}(s=0) = \frac{\wh\f}{2} \frac{\wh \ee^2 (2  + \wh\ee)}{(1+\wh\ee)^3} \qquad
\Rightarrow \qquad
\wh D^{(1)} =  \frac{1}{1 + \frac{\wh \f}{2}\frac{\wh \ee^2 (2  + \wh\ee)}{(1+\wh\ee)^3} } \ ,
\eeq
which in the HS limit reads
\beq\label{eq:D1_HSB}
\wh D^{(1)}_{\rm HS} =  \lim_{\wh\ee \to \io} \wh D^{(1)} (\wh \ee) = \frac{1}{1 + \frac{\wh\f}{2}} .
\eeq
The HS limit of the Laplace transform in Eq.~\eqref{eq:M1SSlB} can be inverted analytically,
yielding
\beq\label{eq:firstHS}
\wt\MM^{(1)}_{\rm HS}(s) =  \frac{\wh\f}{1 + \sqrt{1 + 4s} } \qquad \Rightarrow \qquad
\MM^{(1)}_{\rm HS}(t) = \frac{\wh \f}2 \left[ \frac{ e^{-t /4}}{\sqrt{\p t }} - 
\frac{1}2 \text{erfc} \left( \sqrt{t}/2 \right)  \right] \ ,
\eeq
where $\text{erfc}(x)=1-\erf(x)$ is the complementary error function.
We stress that this is not yet the solution of the dynamical 
Eqs.~(\ref{eq:DYN-SS1-B}-\ref{eq:M-SS1-B}), because we performed only one iteration, 
which corresponds to the lowest order expansion in $\wh\f$.
However, 
Eq.~\eqref{eq:firstHS} provides an important information: for Brownian hard spheres, $\MM(t)\sim \wh\f / \sqrt{4\pi t}$ is divergent for 
small times $t\to 0$. 

Before proceeding, we note that Eqs.~\eqref{eq:D1_HSB} and \eqref{eq:firstHS} are consistent with a series of well-known results:
\begin{itemize}
\item
In the $d\to\io$ limit the memory function also corresponds to the stress autocorrelation~\cite{MKZ15,PUZ20}.
The short-time behavior of
Eq.~\eqref{eq:firstHS} is indeed 
in agreement with the exact short-time behavior of the stress autocorrelation of Brownian hard spheres, as obtained from kinetic theory~\cite{LR94,VDFC97,LCPF09} (see Appendix~\ref{app:MHSshort}).
\item
The short-time divergence is integrable, and it is thus consistent with a liquid phase having finite diffusivity,
given by Eq.~\eqref{eq:D1_HSB}. This result can be interpreted as 
the first-order low-density expansion for Brownian hard spheres, \ie $\widehat D^{(1)}_{\rm HS} \sim 1 - \widehat\varphi/2$. 
It can be compared to previous results from kinetic and linear response 
theory~\cite{HHK82,AF82jcp,LD84jcp}, which predict $D \propto 1 - 2 \varphi$ 
at $d=2,3$. The prediction from linear response can be generalized at 
any $d$ and proven to be consistent with our result when $d \to \io$, 
see Appendix~\ref{app:Dlowphi} for details.
\item
The velocity autocorrelation of hard spheres, and hence the memory function, have long-time tails $\sim t^{-d/2}$~\cite{AF82jcp,AW70}. 
These tails disappear when $d\to\io$ and indeed $\MM^{(1)}_{\rm HS}(t)$ decays exponentially at long times.
Unfortunately, we were unable to generalize the calculation of the memory function given in~\cite{AF82jcp}, which should correspond
to the result in Eq.~\eqref{eq:firstHS}, to arbitrary dimension. 
\end{itemize}

Based on the above results, and on physical intuition,
we conjecture that the Brownian HS dynamics is described by the process
\beq\label{eq:DYN-HSB}
\begin{split}
	\dot h(t)
	&= 1 - \int_{0}^t \de u \,   \MM(t-u) \dot h(u) + \X(t)  \ ,\\
	h(t=0) &= 0 \ , \\
	\langle \X(t) \X(u) \rangle  &= 2 \delta(t-u) + \MM(t-u) \ , \\
	\MM(t)  & =\frac{\wh\f }{2} p(0 , t | 0) \ ,
\end{split}
\eeq
where the process is restricted to $h \geq 0$ with reflecting boundary conditions in $h=0$. We call 
$p(h , t | h_0)$ its propagator, i.e. the probability of starting in $h_0$ at time $t=0$ and arriving in $h$ 
after a time $t$. 
The problem of computing $\MM(t)$ then reduces to the computation of the 
return probability $p(0 , t | 0)$ for the stochastic process defined in Eq.~\eqref{eq:DYN-HSB}.

To support this conjecture, we first note that it is consistent with the exact analysis of 
the first iteration, as given
in Eq.~\eqref{eq:firstHS}.
Indeed, assuming $\MM^{(0)}(t) \equiv 0$, the propagator of a Brownian motion with drift and reflecting barrier in $h=0$ is exactly known~\cite[Appendix 1.16, p. 133]{borodin}, and one finds
\beq \label{eq:p0}
p^{(0)}(0 , t | 0) =
\frac{ e^{-t /4}}{\sqrt{\p t }}
-  \frac12 \text{erfc}(\sqrt{t}/2)
\qquad \Rightarrow \qquad
\MM^{(1)} (t) = \frac{\wh\f}2 p^{(0)}(0 , t|0) \ .
\eeq
A second and more precise argument, valid to all orders in density, is as follows.
We start from the exact formula for $\MM(t)$
in Eq.~\eqref{eq:M-SS1-B}, where 
\beq\label{def_G}
G(h_0,t) = \langle \theta(-h)\rangle_{h_0} = \int_{-\infty}^0 \de h \, p_{\wh\ee}(h, t | h_0) \ ,
\eeq
and $p_{\wh\ee}(h, t|h_0)$ is the propagator in the presence of the SLS potential. We first perform the change of variable $x=\, \wh\ee h_0$ to obtain
\beq\label{change1}
\MM(t) = \frac{\widehat{\varphi}}{2}\int_{-\infty}^0 \de x \, e^{\frac{x}{\wh\ee} + x} \,\wh\ee  \, G\left(\frac{x}{\wh\ee},t\right) \underset{\wh\ee\to\io}\approx  \frac{\widehat{\varphi}}{2} \int_{-\infty}^0 \de x \, e^{x} \,\wh\ee  \, G\left(0,t\right) = \frac{\widehat{\varphi}}{2}  \wh\ee G\left(0,t\right)  \;.
\eeq
We now need to estimate 
$G(0,t)  = \int_{-\io}^0 \de h \, p_{\wh\ee}(h , t| 0)$
in the limit $\wh\ee \to \infty$. In the presence of a strongly repulsive potential $\redv(h)  = - 
\wh\ee\,h$ for negative $h$, it is clear that the trajectories that contribute to this integral start from
$h_0=0$, stay on the positive side up to time $t^-$ and end up on the negative side at some typical value
$h<0$ such that $h \sim {\cal O}(1/\wh\ee)$ or $\wh\ee \, h \sim {\cal O}(1)$. Hence, it is reasonable to
assume that on the negative side we have $p_{\wh\ee}(h , t|0) \sim p_{\wh\ee}(0, t|0) A(\wh\ee h)$. 
Because the motion on the negative side is dominated by the potential repulsion 
for $\wh\ee \gg 1$, the memory and noise terms are thus negligible, 
and it is natural to expect that the function $A(x)$ is independent of time and of density, as it 
only depends on the short time dynamics and on how far the particle can penetrate on the negative 
axis in the presence of $\redv(h)$. Therefore one has
\beq\label{change2}
G(0,t)  = \int_{-\io}^0 \de h \, p_{\wh\ee}(h, t|0) \underset{\wh\ee\to\infty}\approx p_{\wh\ee \to \infty}(0 , t|0) 
\int_{-\io}^0 \de h A(\wh\ee h) \underset{\wh\ee\to\infty}\approx \frac{\AA}{\wh\ee} p_{\wh\ee \to \infty}(0, t|0) \;,
\eeq
where $\AA = \int_{-\io}^0 \de x A(x)$.
In the limit $\wh\ee \to \infty$, because of the repulsive potential, the particle gets immediately reflected on the positive axis as soon as it touches $h=0^-$: therefore $p_{\wh\ee \to \infty}(0,t|0)$ coincides with the propagator $p(0,t|0)$ of the stochastic process in the presence of a reflecting boundary at the origin. Combining Eq.~\eqref{change1} and \eqref{change2}  we obtain
\beq \label{M_final}
\MM_{\rm HS}(t)  =  \frac{\widehat{\varphi}}{2}  \AA\,p(0,t|0) \;, 
\eeq
where we have argued that $\AA$ is time-independent. Its value can thus be fixed by the low-density approximation, for which we can explicitly compute $\AA=1$.

In conclusion, we have shown that the DMFT equations for infinite dimensional 
Brownian hard spheres are those given in Eq.~\eqref{eq:DYN-HSB}.

\subsection{Newtonian dynamics}
\label{ssec:HSN}

In order
to approach the HS limit of the Newtonian dynamical 
Eqs.~(\ref{eq:DYNadim-N}-\ref{eq:Mdefadim-N}), we will proceed in several steps.
First, using a SLS potential, the purely Newtonian dynamics reads
\beq\label{eq:DYN-HSN}
\begin{split}
	\ddot h(t) 
	&= 1 + \wh\ee \th(-h(t))  - \int_{0}^t \de u \,   \MM(t-u) \dot h(u) + \X(t)  \ ,\\
	h(t=0) &= h_0 \ , \\
	\dot h (t=0) &= g_0 \ 	, \\
	\langle \X(t) \X(u) \rangle  &= \MM(t-u) \ ,
\end{split}
\eeq
with the self-consistent condition on the memory function
\beq\label{eq:Mnewt}
\MM(t)   
=\frac{\wh\f}{2}  \int_{-\io}^\io \frac{\de g_0}{\sqrt{2\pi}} e^{-g^2_0/2} \int_{-\io}^0 \de h_0 
\, e^{h_0 + \wh\ee h_0} \wh\ee^2  \langle \th(-h(t))  \rangle_{h_0,g_0} \ .
\eeq
Setting $\MM^{(0)}(t) \equiv 0$, the dynamics in Eq.~\eqref{eq:DYN-HSN} becomes 
deterministic. In the absence of memory, particles follow a piecewise uniformly accelerated 
motion, with acceleration equal to 1 for $h>0$ and to $1+\wee$ for $h<0$. Therefore, 
the only trajectories contributing to $\MM(t)$ in Eq.~\eqref{eq:Mnewt} are those starting 
with $h_0<0$, moving across the negative side until they reach $h=0$, and leaving it 
with positive velocity after the given time $t$. Once a trajectory has left the negative side, 
it will never return to the origin in the absence of noise. 
The condition $h(t)<0$ then becomes equivalent
to $h_0 + g_0 t + (1+\wh\ee) t^2/2 < 0$.
The integral in Eq.~\eqref{eq:Mnewt} can thus 
be analytically computed and gives
\beq\label{eq:Mdelta-SS1}
\MM^{(1)}(t) = 
\frac{\wh\f}{2}  \int_{-\io}^\io \frac{\de g_0}{\sqrt{2\pi}} e^{-g^2_0/2} 
\int_{-\io}^{\min(0,-g_0 t -(1+\wh\ee) t^2/2)} \de h_0 
\, e^{h_0 + \wh\ee h_0} \wh\ee^2 
=\wh\f \frac{\wh\ee^2}{1+\wh\ee} \Th\left( -\frac{1+\wh\ee}{2}t \right) \ ,
\eeq
being $\Th(x) = \int^x_{-\io}\de z \, e^{-z^2/2}/\sqrt{2\pi}$ the normal cumulative distribution function.
In the HS limit $\wh\ee\to\io$, one finds
\beq\label{eq:Mdelta-HS}
\MM^{(1)}_{\mathrm{HS}}(t) = \sqrt{\frac{8}{\pi}} \wh \f \, \delta(t) \ .
\eeq
This result has a clear physical interpretation.
Whenever a trajectory coming from $h>0$ arrives to $h=0$, it undergoes an elastic collision (remember that $h$ is the interparticle gap 
and $h=0$ thus corresponds to two particles being in contact).
In the absence of noise (coming from 
the self-consistent bath of surrounding particles) it is impossible to have multiple collisions, and
the first iteration for the memory kernel thus gives the force-force correlation during the first (and only) 
collision that the particles are possibly undergoing. This correlation decays over 
time scales of order $1/\wee$, and it is delta-peaked in the HS limit as one expects for 
instantaneous collisions.
The same result can be obtained for the SQS potential, 
leading to a memory function which is different from Eq.~\eqref{eq:Mdelta-SS1} but which, as expected, also converges to Eq.~\eqref{eq:Mdelta-HS} in the HS limit.
From the first iteration one can also compute the diffusion coefficient $\wh D^{(1)}$, which 
for Newtonian HS reads
\beq\label{eq:D-HS-1}
\wh D^{(1)}_{\mathrm{HS}} = \left[ \int^{\io}_0 \MM(t) \, \de t \right]^{-1} = 
\sqrt{\frac{\p}{2}} \, \frac{1}{\wh \f} \ .
\eeq
We conclude that, at the lowest order in density, the effect of particle collisions in the HS limit is to add a white noise to the deterministic, Newtonian motion. 

The presence of a delta function in the stress-stress correlation (which coincides with the memory function in infinite dimensions) is well known from kinetic theory~\cite{Du02,BH04,LCPF09}.
The coefficient of the delta function is given by $2/\wh D^{(1)}_{\rm HS}$, as derived above. Note that $\wh D^{(1)}_{\rm HS}$ coincides with
the lowest-order density expansion of the diffusion coefficient for hard spheres at $d=\io$, which can be derived equivalently from the 
Enskog (or Boltzmann) equation~\cite{Hansen} or from kinetic theory~\cite{DEC81,BMD85,MSB01,CCJPZ13},
see Appendix~\ref{app:NewtonianHS1} for details.
The short-time expansion in any dimension also shows that
\beq\label{eq:Mshorttimeexp}
\MM_{\rm HS}(t) = 2 \z_0 \, \delta(t) + B + C t + \cdots \ ,
\qquad
\z_0 = \sqrt{\frac{2}{\pi}} \wh \f = \frac{1}{\wh D^{(1)}_{\mathrm{HS}}} \ ,
\eeq
but the coefficient $B$ vanishes when $d\to\io$,
see Appendix~\ref{app:NewtonianHS1} for a detailed discussion. Hence, the memory function in $d\to\io$
is expected to be 
the sum of a delta function and a regular part that vanishes linearly at short times.

Based on these observations,
we conjecture that the memory function time scales separate for large $\wee$, 
namely
\beq\label{eq:HSNseparation}
\MM(t) \sim 2 \z_0 \wh\ee \MM_{\rm sing}(t \wh\ee) + \MM_{\rm reg}(t) \ ,
\qquad
\int_{-\io}^\io \de t \MM_{\rm sing}(t) = 1 \ ,
\eeq
where the first term becomes a delta function for $\wh\ee\to\io$, and the second 
term is not singular. Because at short times the motion is ballistic and 
dominated by the initial 
velocity, the regular part of the memory function plays no role, and at all orders in density we
expect the result of the first iteration to remain correct, hence 
$\z_0 = \sqrt{2/\p}\, \wh \f$, in agreement with kinetic theory.
On the other hand, for the calculation of the regular part of the memory, we can safely
consider 
$\MM_{\rm sing}(t)$ to be a delta function when $\wh\ee$ is large enough.
We can thus write the effective process as
\beq\label{eq:Newt_interm_1}
\begin{split}
	\ddot h(t) + \z_0 \dot h(t)
	&= 1 + \wh\ee \th(-h(t))  - \int_{0}^t \de u \,   \MM_{\rm reg}(t-u) \dot h(u) + \X(t)  \ ,\\
	\langle \X(t) \X(u) \rangle  &= 2 \z_0 \d(t-u) + \MM_{\rm reg}(t-u) \ ,
\end{split}
\eeq
in which $\z_0$ plays the role of an effective friction coefficient and white noise term. 

At very short times, the motion is given, for large $\wh\ee$, by
\beq\label{eq:Newt_stm}
h(t) = h_0 + g_0 t + \frac12 \wh\ee t^2  + \OO(\sqrt{ \z_0} \, t^{3/2}) \ .
\eeq
Note that the leading correction is a $\OO(\sqrt{ \z_0} t^{3/2})$ term coming from the white noise term in $\ddot h(t)$, which induces a $\dot h(t) \sim (\z_0 t)^{1/2}$ scaling.
Because the typical initial condition is 
$h_0 = -x_0/\wh\ee$ with $x_0 \sim \OO(1)$, the trajectory exits 
from the negative side at a time
$t_{ex}$ with velocity $g_{ex}$, given respectively by
\beq\label{eq:texgex}
t_{ex} = \frac{-g_0 + \sqrt{g_0^2 +2 x_0}}{\wh\ee} \ , 
\qquad
g_{ex} = \sqrt{g_0^2  +2 x_0} \geq 0 \ .
\eeq
Note that the term $\OO(\sqrt{\z_0}\, t^{3/2})$ in Eq.~\eqref{eq:Newt_stm} can be neglected 
because $t \sim 1/\wee$.
The motion for $t<t_{ex}\to 0$ in the region $h<0$ contributes to the delta peak and is not affected by noise, while the motion for $t>t_{ex}$ contributes to the regular
part of the memory function.
For the calculation of the regular part of the memory function, we can thus consider that
trajectories start at $t=0$ in $h_0=0$ with initial postcollisional - i.e. 
positive - velocity $g_{ex}$ given in Eq.~\eqref{eq:texgex}. We thus need to compute the initial distribution of $g_{ex}$. 
The initial distribution of $(h_0,g_0)$ is, for large $\wee$,
\beq
p(h_0,g_0)= \frac{1}{\sqrt{2\pi}} e^{-g^2_0/2}  e^{ \wh\ee h_0} \wh\ee \th(-h_0) \ ,
\eeq
which implies
\beq
\label{eq:pex}
p_{ex}(g_{ex})= \int \de h_0 \de g_0 p(h_0,g_0) \d\left(g_{ex} - \sqrt{g_0^2  -2\wh\ee h_0}\right) = 2 g_{ex}^2 \frac{1}{\sqrt{2\pi}} e^{-g^2_{ex}/2} \ .
\eeq
To summarize, we can now consider $g_{ex} \to g_0$ as the initial velocity, with $h_0=0$, and write the regular part from Eq.~\eqref{eq:Mnewt} as
\beq\label{eq:MregNewt2}
\MM_{\rm reg}(t)   
=\frac{\wh\f}{2}  \int_{0}^\io \de g_{0} p_{ex}(g_0) \lim_{\wh\ee\to\io} \wh\ee \langle \th(-h(t))  \rangle_{h_0=0,g_0}  \ ,
\eeq
where the trajectories evolve according to Eq.~\eqref{eq:Newt_interm_1} in the large $\wee$ limit. Physically, 
while the singular memory contribution is given by the instantaneous collisions, the 
regular part is counting how many trajectories starting after a collision at $t=0$ with postcollisional velocity $g_0$ will come back to collide again at finite time $t>0$ because of 
the noise.

Finally, we need to treat the collision with the barrier that appears at time $t$ in Eq.~\eqref{eq:MregNewt2}.
The trajectory starts in $h_0=0$ with velocity $g_0 \geq 0$ and can undergo multiple collisions in $[0,t]$. We are interested in trajectories
that are negative at time $t$, so let us call $t_1$ the time at which $h(t_1)=0$ for the last time, and $g_1=\dot h(t_1)$ the velocity at that time. 
We can assume that the colliding motion between $t_1$ and $t$ is again dominated by the deterministic part, as in Eq.~\eqref{eq:Newt_stm};
then, this motion is statistically independent from what happened before $t_1$, and we can
write
\beq\label{eq:f-gp}
\langle \th(-h(t))  \rangle_{0,g_0} = \int_{-\io}^0 \de g_1 \int_0^t \de t_1 f(g_1,t_1| g_0 ) P_{\wh\ee}(t-t_1 | g_1) \ ,
\eeq
where 
\begin{itemize}
	\item $f(g_1,t_1| g_0 ) \de t_1 \de g_1$ is the return probability to $h=0$ at a time 
	${\in [t_1,t_1+\de t_1]}$ with (negative) velocity ${\in [g_1,g_1+\de g_1]}$. 
	Note that this probability can also be expressed in terms of the probability $p(g_1,h_1,t | g_0) 
	\de g_1 \de h_1$, which is the probability of finding the particle at time $t$ in a point $\in [h_1, 
	h_1+\de h_1]$ with velocity $\in [g_1,g_1+\de g_1]$, for the limiting process, which is restricted to $h\geq 0$ with a reflecting barrier in 
	$h=0$~\cite{SS05,Bu07}. The relation between $f(g_1,t_1| g_0 )$ and the propagator $p(g_1,h_1,t | g_0)$ reads~\cite{SS05,Bu07}
	\beq\label{eq:f-RTP}
	f(g_1,t_1|g_0) = |g_1| p(g_1,0,t_1 | g_0) \ , \qquad g_1 \leq 0 \ .
	\eeq
	Indeed, every particle that at time $t_1$ is in $h\in [0,|g_1|\de t_1]$ will be found in $h=0$ in a time $t\in[t_1,t_1+\de t_1]$.
	
	\item $P_{\wh\ee}(t-t_1 | g_1)$ is the probability that $h(t)\leq 0$, given that the trajectory has $h(t_1)=0$ and $g(t_1)=g_1$. This quantity
	is immediately computed because
	\beq
	h(t) = g_1 (t-t_1) + \frac{\wh\ee}{2}(t-t_1)^2 \leq 0 
	\qquad
	\Leftrightarrow
	\qquad
	t + \frac{2g_1}{\wh\ee} \leq t_1 \leq t  \ .
	\eeq
	We thus obtain $P_{\wh\ee}(t-t_1 | g_1) = \mathbbm{1}\left(t + \frac{2g_1}{\wh\ee} \leq t_1 \leq t\right)$, where $\mathbbm{1}(\EE)$ is the indicator
	function of event $\EE$.
\end{itemize}
Hence,
\beq\label{eq:pNewt}
\wh\ee \langle \th(-h(t))  \rangle_{0,g_0} = \wh\ee \int_{-\io}^0 \de g_1 \int_{t + \frac{2g_1}{\wh\ee} }^t \de t_1 f(g_1,t_1| g_0 ) 
\approx \int_{-\io}^0 \de g_1 2|g_1|  f(g_1,t| g_0 ) = \int_{-\io}^0 \de g_1 2 g_1^2  p(g_1,0,t| g_0 ) 
\ .
\eeq
Plugging this in Eq.~\eqref{eq:MregNewt2}, we obtain that
\beq\label{eq:MregNewt3}
\MM_{\rm reg}(t)   
=\frac{\wh\f}{2}  \int_{0}^\io \de g_{0} 2 g_{0}^2 \frac{1}{\sqrt{2\pi}} e^{-g^2_{0}/2} \int_{-\io}^0 \de g_1 2 g_1^2 \,  p(g_1,0,t| g_0 )   \ ,
\eeq
which is well defined and regular in the HS limit $\wee \to \io$. 

In conclusion, we have shown that infinite dimensional Newtonian hard spheres are
described by the simple DMFT equations
\beq\label{eq:Newt_interm_2}
\begin{split}
	\ddot h(t) + \z_0 \dot h(t)
	&= 1  - \int_{0}^t \de u \,   \MM_{\rm reg}(t-u) \dot h(u) + \X(t)  \ ,\\
	h(t=0) &= 0 \ , \\
	\dot h(t=0) &= g_0 \ , \\
	\langle \X(t) \X(u) \rangle  &= 2 \z_0 \d(t-u) + \MM_{\rm reg}(t-u) \ ,
\end{split}
\eeq
with a reflecting (elastic) barrier at $h=0$, $\z_0 = \sqrt{2/\p}\, \wh \f$
and the self-consistent condition in Eq.~\eqref{eq:MregNewt3} for the regular part of the memory kernel.
A short time expansion of Eqs.~\eqref{eq:Newt_interm_2} shows that $\MM_{\rm reg}(t) \sim C t$ at short times, hence in Eq.~\eqref{eq:Mshorttimeexp} the coefficient $B=0$ consistently with the kinetic theory expression. Furthermore, we obtain an analytic expression of the coefficient $C = 0.1578 \cdot \wh\f^3$, which to the best of our knowledge has not been obtained via kinetic theory. See Appendix~\ref{app:NewtonianHS2} for details.

\subsection{Plateau equation}

We now check that the DMFT equations for hard spheres give the correct equation for the plateau of the memory function in the glass phase, as derived from a thermodynamic analysis~\cite{PUZ20}. With the same decomposition of $\MM(t)$ and $\Xi(t)$ as in section~\ref{ssec:plateau}, we obtain
\beq
\LL[h] = 1 - \MM_\io h + \Xi_\io = -w'(h) \ ,
\eeq
recalling that for hard spheres $h_0=0$, and that here we are using dimensionless equations. Because there is a reflecting barrier in $h=0$, the long time distribution is
\beq
p(h|\Xi_\io) = \frac{e^{- w(h)}\th(h)}
{\int_0^\io \de z e^{-w(z)}} \ ,
\qquad w(h) = -h + \frac{\MM_\io}2 h^2 -\Xi_\io h \ .
\eeq
In the Brownian case, the long-time limit of the memory function is given by
\beq\label{eq:MplateauHS}
\MM_\io = \frac{\wh\f}2 p(0,t\to\io|0) = 
\frac{\wh\f}2 \int \de \Xi_\io p(\Xi_\io) p(0|\Xi_\io)=
\frac{\wh\f}2 \int \de \Xi_\io p(\Xi_\io) 
\frac{1}
{\int_0^\io \de h \, e^{h - \frac{\MM_\io}2 h^2 +\Xi_\io h}} \ ,
\eeq
which provides a simple self-consistent equation for $\MM_\io$, recalling that $p(\Xi_\io)$ is a Gaussian with zero mean and variance $\MM_\io$.
In the Newtonian case, the distribution of $p(g,h,t|g_0)$ at long times is given by $p(h|\Xi_\io) p(g)$, where $p(g)$ is a centered unit Gaussian. Plugging this in Eq.~\eqref{eq:MregNewt3}, the integrals over $g_0$ and $g_1$ evaluate to one, and we obtain the same result as in the Brownian case. 
Note that Eq.~\eqref{eq:MplateauHS} is indeed the hard sphere limit of Eq.~\eqref{eq:Mplateau}, as one can check explicitly using the SLS potential in Eq.~\eqref{eq:Mplateau} and taking the limit $\wee\to\io$. The plateau Eq.~\eqref{eq:MplateauHS} admits only the liquid solution $\MM_\io=0$ for $\wh\f<4.8067\ldots$ and admits a non-trivial solution $\MM_\io>0$ for $\wh\f>4.8067\ldots$~\cite{PUZ20}.

\section{Numerical algorithms}
\label{sec:num}

In this section, we give some details on how to compute $\MM(t)$ through the numerical integration 
of the stochastic differential equations of DMFT. The 
numerical scheme is close to that used in~\cite{RBBC19}.

\subsection{Solution scheme}

In order to compute $\MM(t)$ numerically, one needs to solve the 
stochastic process in Eq.~\eqref{eq:DYN}, in proper units as discussed in section~\ref{ssec:adim}. 
For numerical convenience, we used the quadratic (harmonic) 
soft sphere potential (SQS), $\redv(h) = \wh \ee h^2 
\theta(-h)/2$, where $\theta(x)$ denotes the Heaviside step function. 
This potential reduces to the hard sphere potential when $\wh \ee \to \io$,
it has a continuous derivative in $h=0$, and it grows quickly when $h\to-\io$. 
This choice allows one to restrict the integration over $h_0$ to a small 
region, because only the terms with $h_0<0$ contribute to 
Eq.~\eqref{eq:Mdef}, and those are weighted by a
Gaussian-shaped distribution, namely
\beq\label{eq:M_HSS}
\MM(t) = \frac{\wh \f}{2} \wh\ee^2 \int^{\io}_{-\io} \frac{\de g_0}{\sqrt{2\pi}} e^{-g^2_0/2} \int^0_{-\io} \de h_0 \,e^{h_0-\wh\ee h^2_0/2} h_0 \left\langle h(t) \theta(-h(t)) \right\rangle_{h_0, g_0} \ .
\eeq
The integrals in Eq.~\eqref{eq:M_HSS} are numerically computed by running trajectories 
starting in $h_\min < h_0 < 0$, with a typical cut-off $h_\min = - 
5/\sqrt{\wh\ee}$, in such a way that the contributions coming from $h_0 < h_\min$ are bounded by 
$e^{-\wh\ee h_\min^2/2} < 4 \times 10^{-6}$, and drawing a random Gaussian initial velocity $g_0$ 
in the Newtonian case.
Conversely, in the hard sphere case one needs to compute $\MM(t)$ 
through Eqs.~\eqref{eq:DYN-HSB} or~\eqref{eq:MregNewt3}. Therefore, 
all the trajectories start at $h_0=0$ (postcollisional condition), 
with a $g_0$ distributed according to Eq.~\eqref{eq:pex} in the Newtonian 
case.

Being at equilibrium, the system is time-translational invariant (TTI) 
and this property can be exploited to generate the correlated noise; 
indeed, if the noise correlation $\KK(t)$ is TTI, the associated noise will 
be delta-correlated in frequency, namely $\left\langle  
\Xi(\om) \Xi^*(\om') \right\rangle =  \KK(\om)
\d(\om -\om')$. One can generate independently the noise in Fourier space, 
$\Xi(\om)$, with zero mean and variance $\KK(\om)$,
and its inverse Fourier transform 
thus generates a correlated Gaussian noise $\Xi(t)$ with the desired 
correlation $\left\langle \Xi(t) \Xi(t') \right\rangle = \KK(t-t')$. Note that, depending on the context, the white noise part can be generated independently and $\KK(t)=\MM(t)$,
or it can be included in the Fourier transform
and $\KK(t) = 2 \d(t) + \MM(t)$.
The noise is generated at the beginning and it is then injected in the equation of motion for $h(t)$, which is then integrated to obtain the trajectory.

The iterations stop when a convergence criterion is reached: in our 
algorithm, we typically required that the rescaled viscosity $\wh \eta_s
\propto \int^{\io}_0 \MM(t) \, \de t $ does not change significantly between 
iterations $i$ and $i+1$, namely $\vert (\wh\eta^{i+1}_s-\wh\eta_s^i)/\wh\eta_s^i \vert < 
\delta $, where $\delta$ is a small parameter fixing the relative error. Obviously, many other convergence criteria can be chosen and 
some others have been tested to ensure that the results do not 
depend significantly on this arbitrary choice. We typically report numerical solutions with $\d = 10^{-3}$, which is a compromise 
between a satisfyingly low relative error and a reasonable number of 
trajectories being needed to reduce fluctuations. One can also achieve smaller 
values of $\d$ by increasing the number of trajectories while approaching
convergence, which however increases the convergence time.

\subsection{Convergence algorithms}
\label{ssec:convalg}

The convergence algorithm discussed in section~\ref{ssec:iterative} has been implemented on a fixed 
time grid, which yields a solution for $\MM_n = \MM (n \D t)$ with $0 \leq t < t_\max = N_S \D t$, being $N_S$ the total number of time steps. This is the most straightforward way to approach the 
problem, but the maximum value of $N_S$ is constrained by the computation of the retarded 
friction in Eq.~\eqref{eq:DYN}, which has a time complexity scaling as $N^2_S$. 

More sophisticated algorithms can be developed to improve the computational efficiency
and at the same time check the validity of the results. The main 
observation is that the DMFT equations are causal, \ie the solution $\MM_n$ 
is independent on future times $m \D t > n \D t$. One can then compute 
a solution up to a final time $t_1$, then fix the value of $\MM(t)$ for 
$0\leq t<t_1$ and extend the trajectories up to $t_2>t_1$, compute $\MM(t)$ 
in the new time window and so on. In our study, we also developed two 
algorithms exploiting causality: a step-by-step algorithm and a 
decimation algorithm.

The step-by-step algorithm computes recursively $\MM_n$ starting 
from $\MM_0$ -which is analytically known from Eq.~\eqref{eq:Mdef}-, keeping track 
of the noise realizations for $m = 0,1,\ldots,n-1$ and drawing the 
noise at the $n$-th step conditioned to the previous ones. This method does not 
require any convergence criterion for the global memory function, but only on the fluctuations of the new memory value being 
computed. However, the generation of the trajectories cannot be 
performed in Fourier space as explained above because of the bias 
introduced by the past realizations, and one needs to invert a correlation
matrix at any time step. 
This method suffers, however, a serious limitation when applied to equilibrium dynamics. In fact, 
computing the matrix $\MM_{m,n}$ using the 
TTI assumption as in Eq.~\eqref{eq:Mdef}, which gives $\MM_{m,n} = \MM_{0,\vert m-n \vert}$, preserves the positivity of $\MM_{m,n}$ only in the limit of an infinite number of trajectories. With a finite number of trajectories,
$\MM_{m,n} = \MM_{0,\vert m-n \vert}$
can have negative eigenvalues because 
of statistical fluctuations in the numerical solution. 
This issue would not be present if the memory function was
computed without assuming TTI, because in that case it is easy to show that 
$\MM_{m,n} \propto \langle \redv'(h(m\Delta t)) \redv'(h(n\Delta t))  \rangle$ (where the average is over trajectories with the proper initial conditions) is a positive-definite matrix for any number of simulated trajectories. 
However, calculating the memory in this way is more difficult, because it receives contributions from 
trajectories starting in any $h_0 >0$, which introduces the non-trivial problem of finding an upper cutoff 
$h_0^\max$ for the $h_0$ integral.
Therefore, we only used this method to simulate short-time trajectories; the results we found agree with the 
fixed time grid method. Note that in future non-equilibrium studies~\cite{AMZ18,AMZ19} the TTI hypothesis will have to be relaxed anyway, and this method is then more interesting than the fixed-grid method.

Another algorithm we considered is the so-called decimation 
algorithm~\cite{KL01}: because of the causality, one can compute $\MM_n$ on a 
fixed grid up to a final time $t_1$ with time steps $\D t_1$; then, 
one can double the time step and the final time, defining 
$\D t_2 = 2 \D t_1$ and $t_2 = 2 t_1$. The memory function is 
computed iteratively as on the fixed time grid, but keeping fixed the 
part of $\MM_n$ corresponding to $n \D t_2 < t_1$, using an exponential fit as 
initial condition on the second half, and so on, iteratively.
This method has the great advantage of providing a higher resolution at short times, 
\ie starting with a $\D t_1 \sim 10^{-5}$, and reaching final times 
$t_\max \sim 100$ with an increased efficiency with respect to the fixed 
time grid algorithm; however, at variance with the decimation algorithm 
used in numerical solutions of MCT-like equations~\cite{KL01}, here we cannot increase $\D t$ indefinitely because, while the memory function decays slowly at large times, the individual trajectories $h(t)$ still fluctuate wildly.
Because we do not know the explicit time propagator of the probability density of $h$, the 
time step $\D t$ must be kept small enough to ensure a correct integration of the equation of motion for $h(t)$. Furthermore, 
the main advantage of decimation algorithms is the speedup of the 
computation of the memory integral through the decomposition in a slow and a fast part~\cite{KL01}; unfortunately, this 
is not possible in our model because, once again, trajectories $h(t)$ cannot be split in a slow and fast decay, 
as it is usually done for correlation functions.
Another minor problem of the decimation algorithm is the appearance 
of discontinuities in the solution for $\MM(t)$ at the boundary of each 
grid, indicating an imperfect matching of numerical solutions when changing the time step $\D t$.

For all these reasons, most of the numerical results shown in section~\ref{sec:results} are obtained via the
fixed grid algorithm.
In the specific case of Brownian hard spheres, however, we will show that because of the divergence of $\MM(t)$ at short times, 
a decimation algorithm provides better solutions than the fixed grid algorithm.

\subsection{Dynamical equation in discrete time}
\label{ssec:discretization}

We now discuss how the DMFT Eqs.~\eqref{eq:DYN} can be 
discretized over a fixed time grid with time steps $\D t$. 
We will use both the SQS potential 
and a HS potential, in the Brownian and Newtonian cases.

\subsubsection{Soft spheres}
\label{ssec:SS-disc}

We obtained $N_T = N_H \times N_P$ trajectories determined by 
Eq.~\eqref{eq:DYN}, integrating $N_P$ independent noise realizations for each of 
$N_H$ values of $h_0$ chosen on a uniform grid between $h_\min$ 
and 0. Because the correlation of the noise $\Xi(t)$ has both a white and a 
colored contribution, we splitted it into two independent Gaussian noises
$\Xi(t) = \xi(t) + \chi(t)$, having respectively $\left\langle \xi(t) 
\xi(u) \right\rangle = 2 \d(t-u)$ and $\left\langle \chi(t) \chi(u) 
\right\rangle = \MM(t-u)$. 

The Brownian dynamical Eq.~\eqref{eq:DYNadim-B}, discretized within It\^o calculus by 
the Euler-Maruyama method, reads
\beq\label{eq:numSSB}
\begin{split}
	h_{n+1} &= h_n + \left( 1 - \wh\ee \, h_n \, \theta(-h_n) - I_n + \chi_n \right) \Delta t + \sqrt{2\Delta t} \, \xi_n = h_n + \Delta h_n \: , \\
	I_n &= \sum^{n-1}_{m=0} \MM_{n-m} \Delta h_m \: .
\end{split}
\eeq
Here $\xi_n$ and $\chi_n$ are independent random 
variables with zero 
average and correlations $\left\langle \xi_m \xi_n 
\right\rangle = \d_{m,n}$ and $\left\langle \chi_m \chi_n \right\rangle = 
\MM_{\vert m-n \vert}$. The memory kernel $\MM_n$ is updated after one 
iteration through the numerical evaluation of Eq.~\eqref{eq:M_HSS}, namely
\beq
\label{eq:Mnum-SS}
\MM_n = \frac{\wh\f}{2} \wee^2 \sum_{i=1}^{N_H} e^{h_{i} - \wee h^2_{i} /2} \, h_{i} \, \D h \frac{1}{N_P} \sum_{p=1}^{N_P} h_n^{(p,i)} \, \theta \left( - h_n^{(p,i)} \right) ,
\eeq
where $\D h = |h_{\min}|/N_H $ and $h_i = h_\min + (i-1/2)\D h$, while 
$h_n^{(p,i)}$ is the $p$-th stochastic trajectory realized starting from 
$h_i$.
The iterations are repeated until 
convergence, as explained in section~\ref{ssec:convalg}.

Within Newtonian dynamics, the second-order Eq.~\eqref{eq:DYNadim-N}
can be written as a system of two coupled first-order equations for 
$h(t)$ and $g(t) = \dot h(t)$. The scheme is the same as for Brownian 
dynamics, except that one must draw an initial velocity $g_0$ from 
a normal distribution with unit variance, and we discretize 
the motion through a stochastic Verlet algorithm, which reads
\beq\label{eq:numSSN}
\begin{split}
    f_n &= 1 - \wh\ee \, h_n \, \theta(-h_n) - I_n + \chi_n \: , \\
    \wt g_n &= g_n + \frac{1}{2} f_n \D t \: , \\
    h_{n+1} &= h_n + \wt g_n \D t \: , \\
    g_{n+1} &= \wt g_n + \frac{1}{2} f_{n+1} \D t \: , \\
	I_n &= \sum^{n-1}_{m=0} \MM_{n-m} \Delta h_m \: ,
\end{split}
\eeq
with $\la \chi_m \chi_n \ra = \MM_{|m-n|}$. 
The numerical evaluation of $\MM_n$ is again obtained via 
Eq.~\eqref{eq:Mnum-SS}, with an additional average over the random initial velocity, randomly drawn from a unit centered Gaussian.

\subsubsection{Hard spheres}
\label{ssec:HS-disc}

In order
to integrate the HS dynamics, one needs to implement an appropriate 
reflecting boundary condition at $h=0$. Contrary to the soft spheres case, 
all the trajectories that contribute to $\MM(t)$ start from $h=0$ and eventually 
collide with the 
boundary between $t$ and $t + \D t$. Therefore one has $N_T=N_P$ because 
$N_H=1$. Another problem is given by the interpretation of $\MM_0$, 
which is infinite for hard spheres, but that must be set equal to a 
physically meaningful value in the discretization.

For Brownian dynamics, we include the white component of the noise 
in the definition of the noise kernel. Therefore, we have 
$ \langle \Xi_m \Xi_n \rangle = (2 / \D t ) \, \d_{m,n} + 
\MM_{\vert m-n \vert} $, we fix $\MM_0 = 2 / \D t$, 
and we only update the memory for $n>0$. 
The infinite value of $\MM_0$ 
and the short-time singularity $\MM(t)\propto t^{-1/2}$
could in principle 
also affect the memory 
integral $I_n$; however, 
when we approximate the retarded friction with a rectangle sum as in 
the last lines of Eqs.~\eqref{eq:numSSB}, \eqref{eq:numSSN}, we then 
make an error of order $\sqrt{\D t}$ in the integration, 
which becomes an error of order $\D t^{3/2}$ in the dynamical equation.
For this reason, we neglect it as a first approximation.
The Brownian dynamics is thus discretized with the rule
\beq\label{eq:numHSB}
\begin{split}
	h_{n+1} &= \vert h_n + \left( 1 - I_n + \Xi_n \right) \Delta t \vert = h_n + \Delta h_n \: , \\
	I_n &= \sum^{n-1}_{m=0} \MM_{n-m} \Delta h_m \: ,
\end{split}
\eeq
where the modulus enforces the reflecting boundary condition in $h=0$.
A collision occurs between $n \D t$ and $(n+1) \D t$ if the argument 
of the modulus in Eq.~\eqref{eq:numHSB} is negative; in this case, 
we add a count to the collisional probability at time $n$, which we call
$P_n$. The latter quantity is related to the continuous-time return 
probability $p(0,t|0)$ defined in Eq.~\eqref{eq:DYN-HSB} observing that, for 
$\D t \ll 1$, a collision occur if $h_n + \sqrt{2\D t} \, \xi_n < 0$ 
(considering only the white part of the noise which dominates the short time motion).
Therefore
\beq
\label{eq:probHS}
\begin{split}
	P_n &\simeq \int^\io_0 \de h \, p(h,t=n\D t \vert 0 ) \int^{\io}_{-\io} \frac{\de \xi}{\sqrt{2\p}} 
	e^{-\xi^2/2} \, \th \left(-h-\sqrt{2\D t} \, \xi\right) \\
	& = \int^\io_0 \de h \, p(h,t \vert 0) \, \Th \left( -h/\sqrt{2 \D t} \right) \simeq \sqrt{\frac{\D t}{\p}} p(0,t|0) \ ,
\end{split}
\eeq
being $\Th(x) = (2 \p)^{-1/2} \int^x_{-\io} \de z \, e^{-z^2/2}$ the 
normal cumulative distribution function and having also assumed that 
$p(h,t \vert 0)$ is continuous for $h \to 0^+$. 
The numerical estimate of $P_n$ is given by 
\beq \label{eq:Pnum}
P_n = \frac{1}{N_P} \sum^{N_P}_{p=1} C^{(p)}_n
\eeq
being $C^{(p)}_n = 1$ if the $p$-th trajectory collides between 
$n \D t$ and $(n+1) \D t$, and 0 otherwise.
Therefore, the memory kernel is computed at every iteration through the 
collisional probability $P_n$ with the following expression, derived 
from Eq.~\eqref{eq:DYN-HSB}:
\beq \label{eq:Mnum-HSB}
\MM_n = \frac{\wh \f}{2} \sqrt{\frac{\pi}{\D t}} P_n .
\eeq

The algorithm for Newtonian hard spheres is similar. In this case, we only need to compute the regular part of the memory function as defined in 
Eq.~\eqref{eq:MregNewt3}, while we already know that there is a delta 
peak at $t=0$ with amplitude $\z_0 = \sqrt{2/\p} \wh \f$ as derived 
in section~\ref{ssec:HSN}. We thus set $\MM_0 = \z_0 / \D t$ analogously 
to the Brownian case, and we simulate the dynamics with a single noise 
term $\Xi_n$ having correlation $\langle \Xi_m \Xi_n \rangle = 
(\z_0 / \D t) \, \d_{m,n} + \MM_{\vert m-n \vert}$. In this case, we 
discretize Eq.~\eqref{eq:Newt_interm_2} with a stochastic Verlet 
algorithm, which reads
\beq\label{eq:numHSN}
\begin{split}
    f_n (g_n) &= 1 - \z_0 \, g_n  - I_n + \Xi_n \: , \\
    \wt g_n &= g_n + \frac{1}{2} f_n(g_n) \D t \: , \\
    h_{n+1} &= \vert h_n + \wt g_n \D t \vert \: , \\
    g_{n+1} &=  
    \begin{cases} \wt g_n + \frac{1}{2} f_n(\wt g_n) \D t 
    &\text{if} \quad
    h_n + \wt g_n \D t >0 \: , \\
    - g_n  &\text{if} \quad h_n + \wt g_n \D t < 0 \ ,
    \end{cases} 
     \\
	I_n &= \sum^{n-1}_{m=0} \MM_{n-m} \Delta h_m \: .
\end{split}
\eeq
Note that in case of a collision we simply reflect both
the final position of the particle -as in the Brownian case- and 
the velocity, $g_{n+1} = - g_n$, \ie in that case we do not use the Verlet construction. Because collisions happen at times that are separated by a $\OO(1)$ interval, errors during collisions do not accumulate and it is safe to discard higher order corrections.

When a collision occurs, the collisional probability is updated
as in the Brownian case; however, because in the Newtonian dynamics the 
memory function is given by Eq.~\eqref{eq:MregNewt3}, the collisions 
must be weighted. From Eq.~\eqref{eq:f-gp}, we know that the 
probability to collide between $t$ and $t+\D t$ with velocity $g_1$ 
having started at $h=0$ with velocity $g_0$ is equivalent to 
$f(g_1,t|g_0)$. Thus, from the third equivalence of Eq.~\eqref{eq:pNewt} 
we need to measure the average of $2 \vert g_1 \vert C_n$, being 
$C_n$ the collisional event defined in the Brownian case and $g_1$ the 
precollisional velocity. The memory function is then given by
\beq\label{eq:MnumHSN}
\MM_n = \frac{\wh \f}{2} \frac{1}{N_P \, \D t} \sum^{N_P}_{p=1} 
2 \, \vert g^{(p)}_{n-1} \vert  \, C^{(p)}_n \ ,
\eeq
having started the trajectories in $h=0$ with an initial velocity $g_0$ 
distributed as $P(g_0) = \sqrt{2/\p}\, g^2_0\, \exp(-g^2_0/2)\, \th(g_0)$.

\begin{figure}[t]
    \centering
	\includegraphics[width=0.47\textwidth]{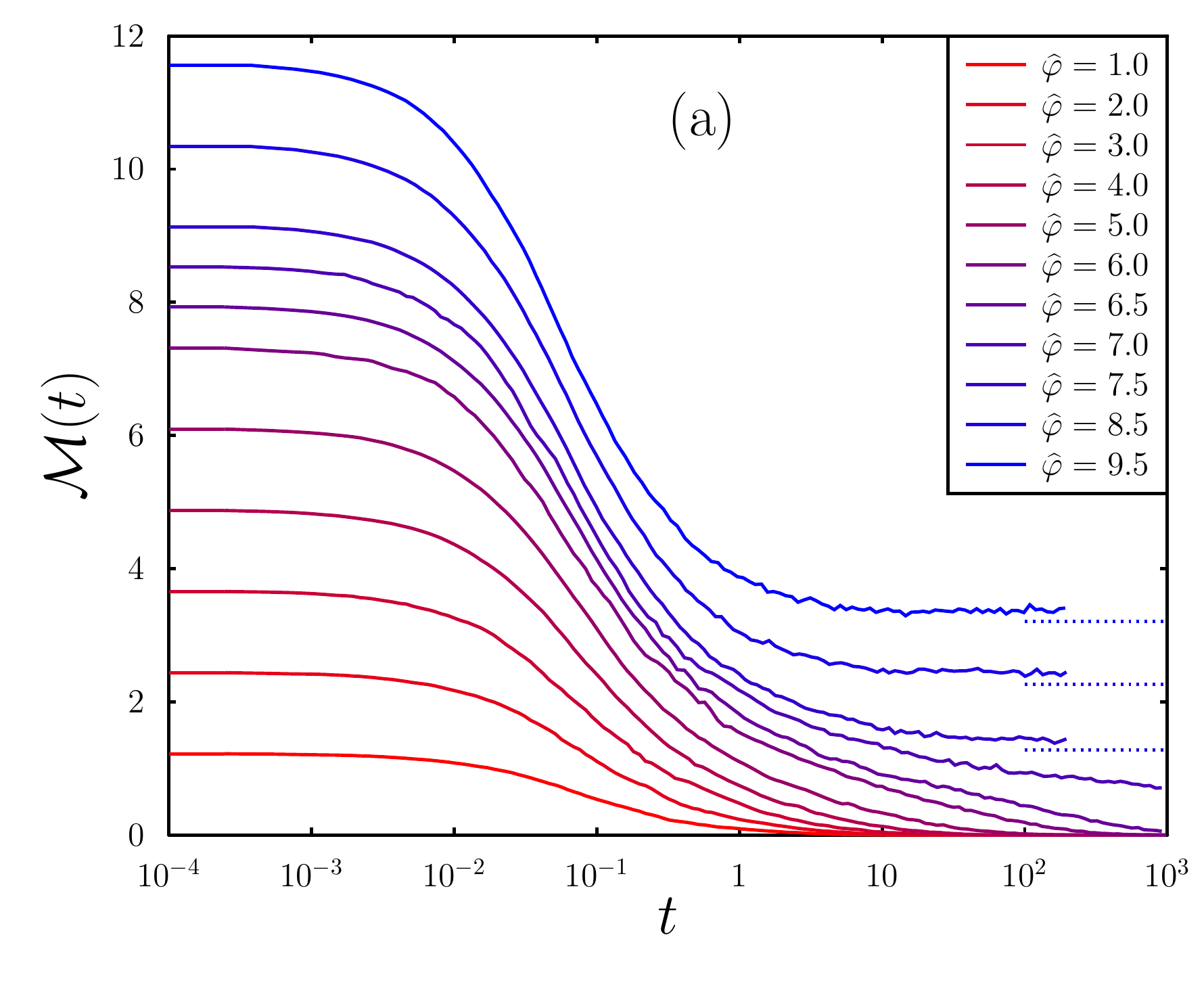}
	~
	\includegraphics[width=0.47\textwidth]{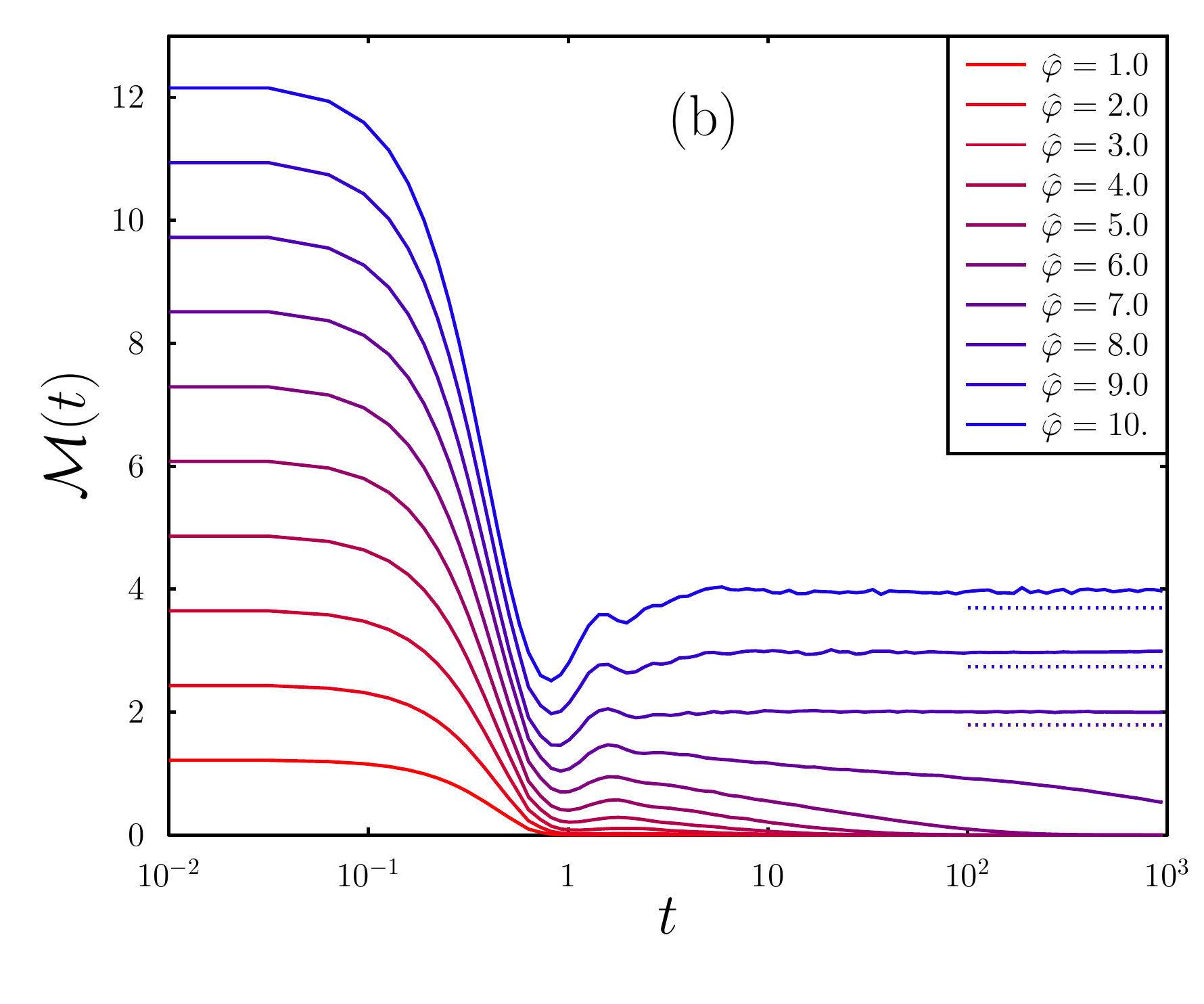}
	~
	\includegraphics[width=0.47\textwidth]{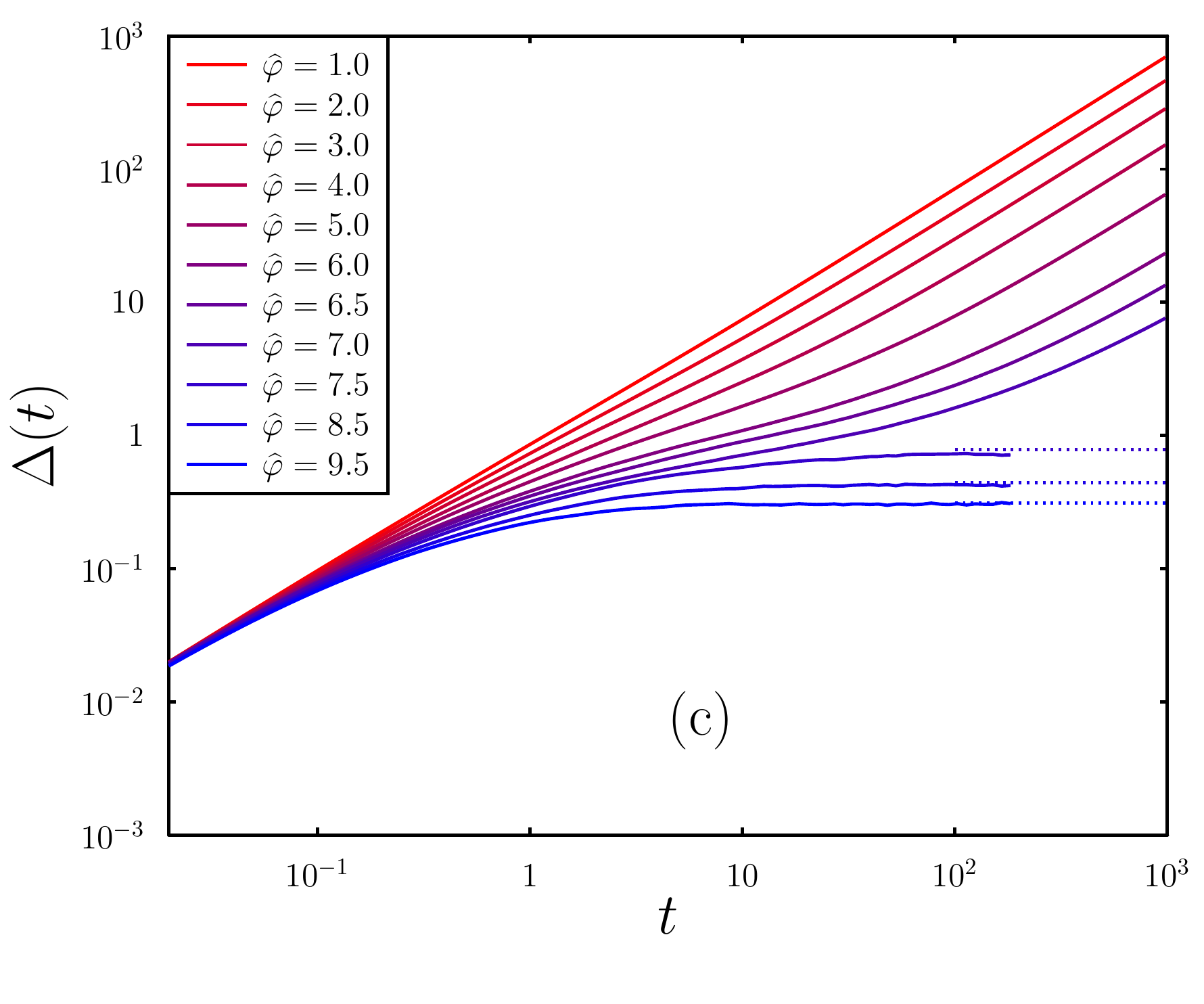}
	~
	\includegraphics[width=0.47\textwidth]{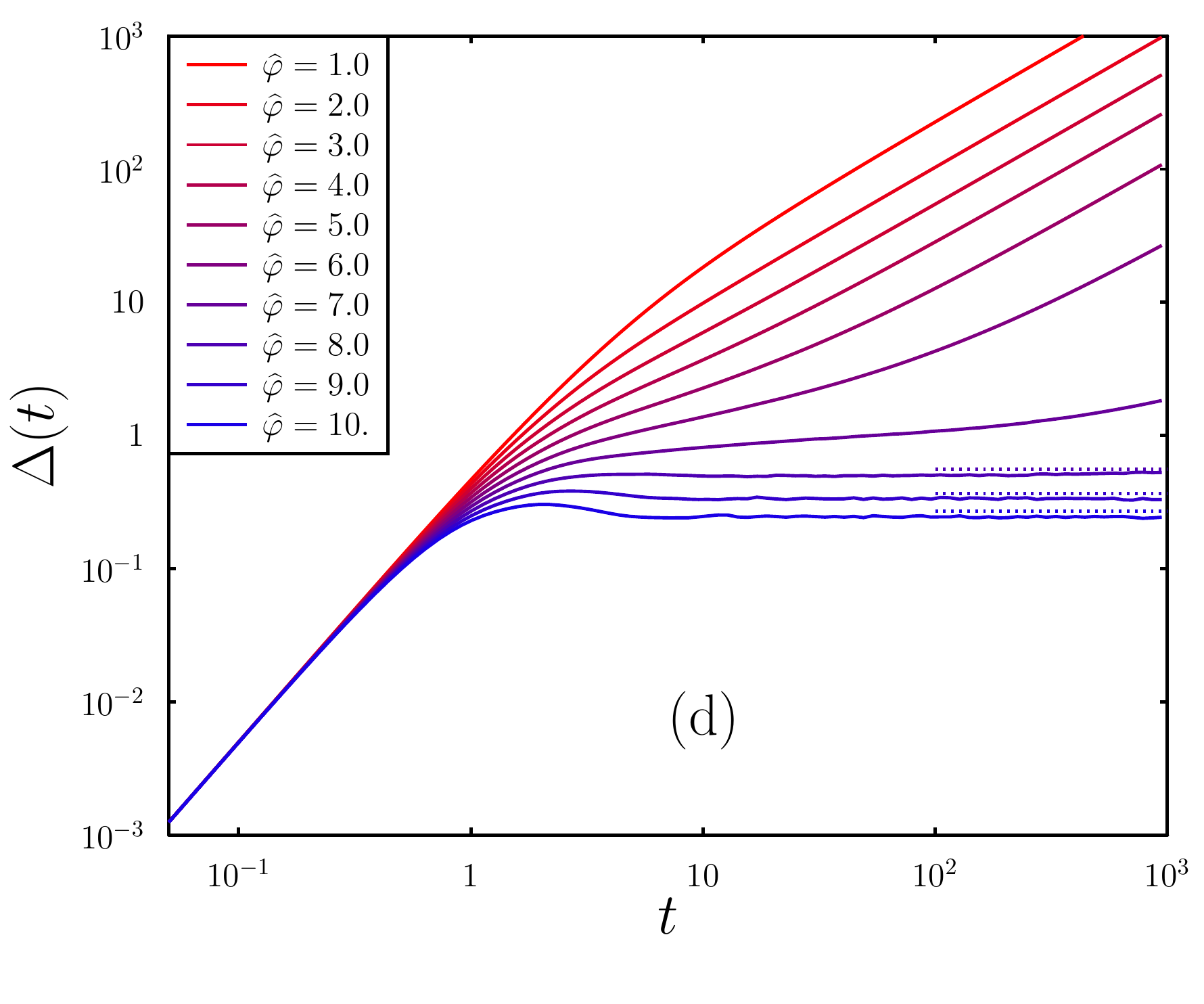}
	\caption{
	Numerical solution of the DMFT equations for the quadratic soft sphere potential at $\wee=10$ and several densities~$\wh\f$.
	Top panels: memory kernels $\MM(t)$ for Brownian (a) and Newtonian (b) dynamics.
	The memory functions 
	in the liquid phase $\wh\f < \wfd \simeq 7.2458$ decays over a time scale that 
	increases upon approaching the dynamical transition. In the dynamically arrested phase $\wh\f > \wfd$,
	a plateau emerges and $\MM(t) \to \MM_\io >0$ for long times. The value of 
	$\MM_\io(\wh\f)$ obtained from the plateau equation is plotted as a dashed lines with the same color as the corresponding density. Bottom 
	panels: mean square displacement $\D(t)$ for Brownian (c) and Newtonian (d) dynamics. The long-time limit 
	$\D_\io = T^2 / \MM_\io$ in the dynamically arrested phase is plotted as a dashed line.
	}
	\label{fig:SS}
\end{figure}

\section{Results}
\label{sec:results}

In this section, we present results for $\MM(t)$ and $\D(t)$
for soft and hard spheres, in both cases for Brownian and Newtonian dynamics. For soft spheres, we always use the quadratic potential, SQS with $\wee=10$, unless otherwise specified. From the analysis of the plateau equations derived in section~\ref{ssec:plateau}, the dynamical glass transition happens at density $\wh\f_{\rm d} \simeq 7.2458\ldots$ in this case. For hard spheres, the dynamical glass transition is at $\wh\f^{\rm HS}_{\rm d}=4.8067\ldots$. Note that the dynamical glass transition line for SQS scales
as $\wh\f_{\rm d} - \wh\f_{\rm d}^{\rm HS} \sim 6.13/\sqrt{\wee}$ for $\wee\to\io$~\cite{SBZ18}, which explains why the dynamical transition of SQS is quite distinct from that of hard spheres even at rather large $\wee$.

\subsection{Soft spheres}
\label{ssec:results-SS}

\begin{figure}[t]
	\centering
	\includegraphics[width=0.47\textwidth]{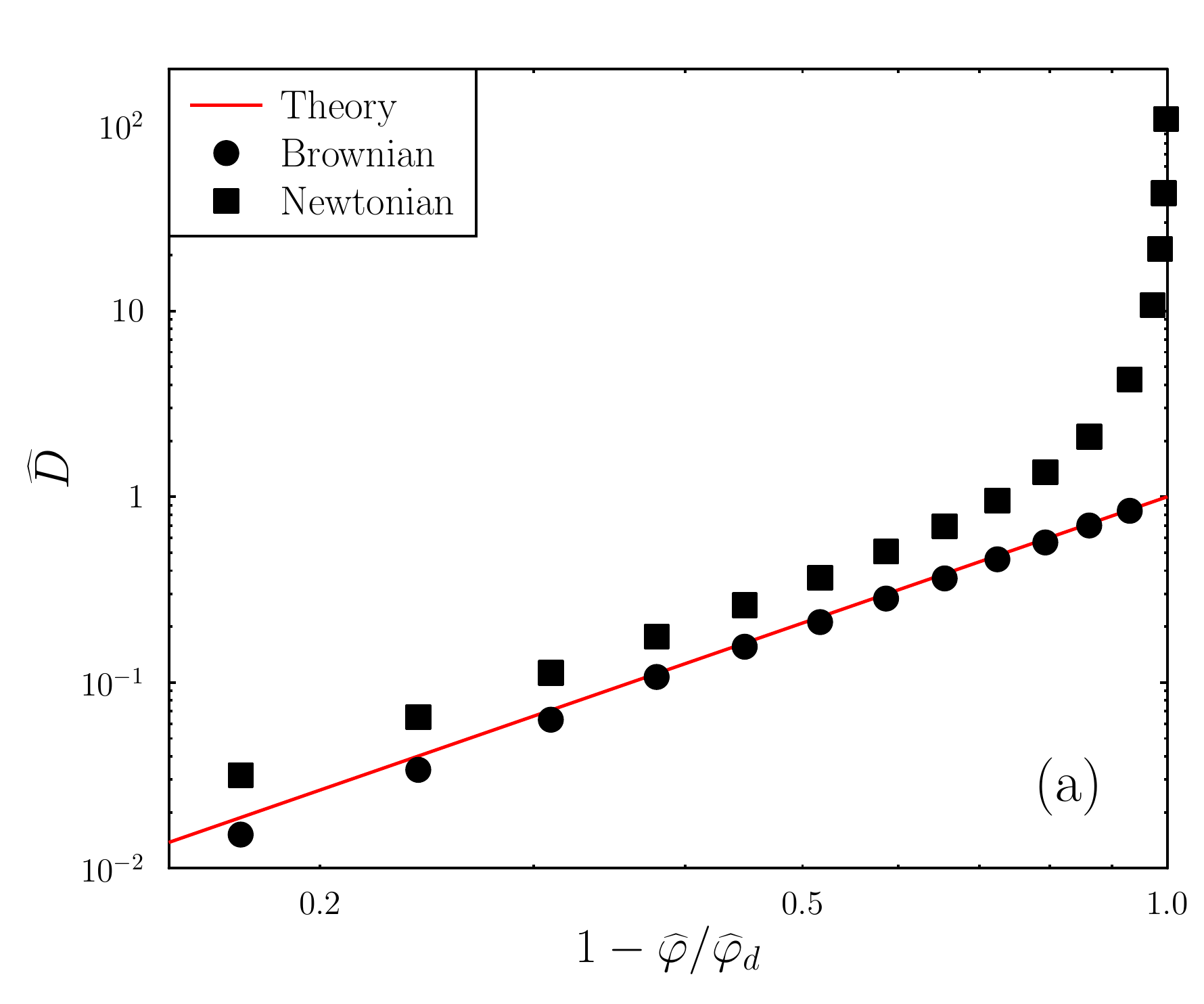}
	~
	\includegraphics[width=0.47\textwidth]{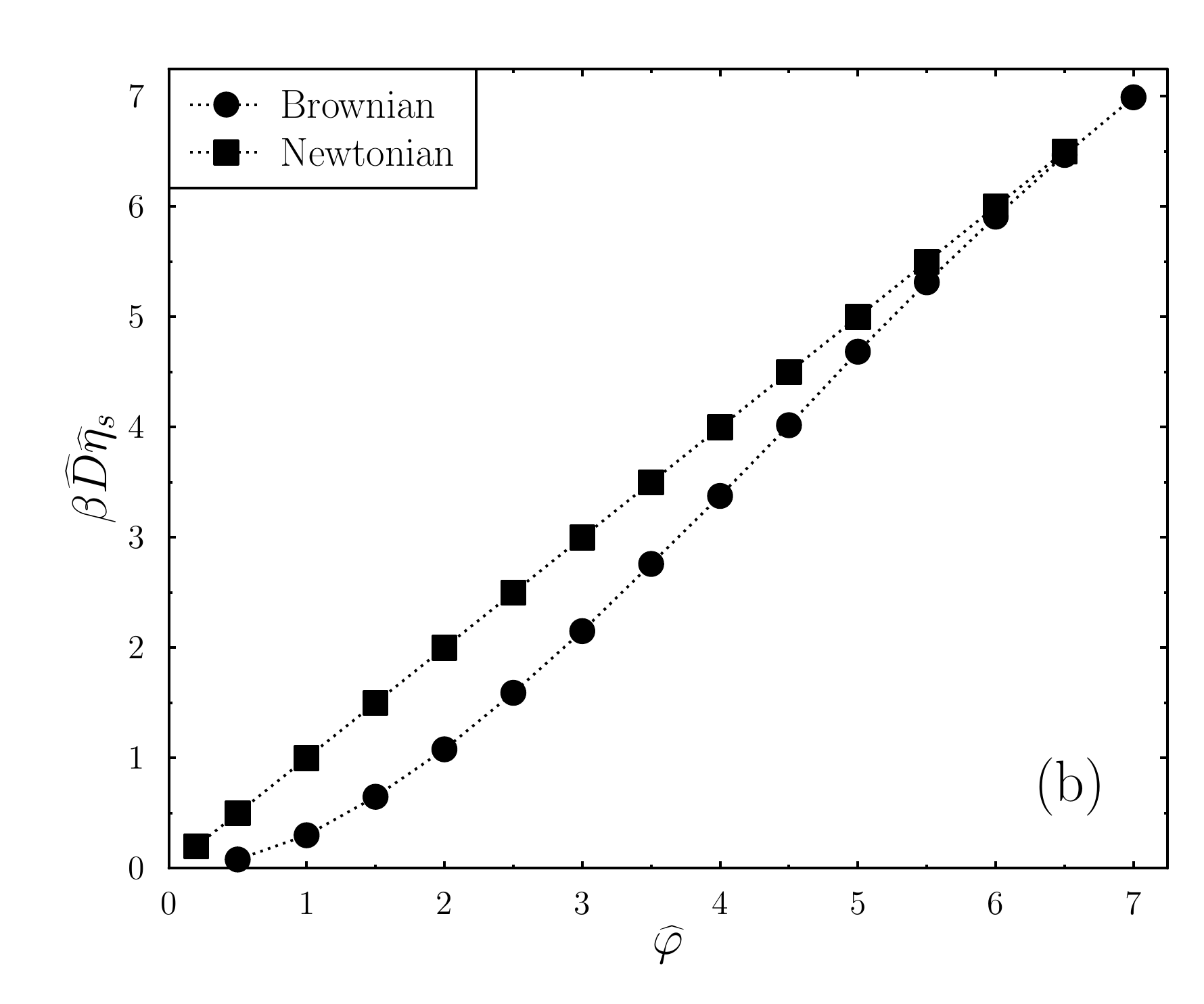}
	\caption{
	Numerical solution of the DMFT equations for the quadratic soft sphere potential at $\wee=10$ and several densities $\wh\f$.
	(a)~Critical scaling of the diffusivity $\wh D \sim (\wh\f_d - \wh\f)^\g$
	    for $\wh\f \to \wh\f_d^-$, for Brownian (circles) and Newtonian (squares) dynamics. 
	    The line is the analytical prediction with $\wfd=7.2458$ and $\g \simeq 2.25844$; the height 
	    of the line is adjusted to the plot.
	    (b)~Stokes-Einstein relation for Brownian and Newtonian dynamics, same key as left panel.
	    }
	\label{fig:diffusivity-SS}
\end{figure}

Results for 
$\MM(t)$ obtained in Brownian and Newtonian dynamics are shown in 
Figs.~\ref{fig:SS}a,~\ref{fig:SS}b, respectively.
The characteristic decay time of $\MM(t)$ increases 
upon increasing 
density $\wh \f$, corresponding to a dynamical slowing down, which becomes heavily 
pronounced for $\wh\f=7.0$. For $\wh\f \geq 7.50$, the memory exhibits a 
plateau as expected in the dynamically arrested phase.
The numerical values of the memory can be compared to the analytical result for the plateau obtained from Eq.~\eqref{eq:Mplateau}.
The comparison is shown in Figs.~\ref{fig:SS}a,~\ref{fig:SS}b, 
with a fair agreement between analytical and numerical results.
Note that while in the Brownian case $\MM(t)$ is a monotonically decreasing function of $t$, in the Newtonian case we observe 
characteristic oscillations at intermediate times.

From the knowledge of $\MM(t)$, the mean square displacement (MSD) can be easily computed through 
Eq.~\eqref{eq:MSD}: the numerical results for the MSD versus time are 
shown in Figs.~\ref{fig:SS}c,~\ref{fig:SS}d. 
For $\wh\f < \wfd$, one finds the free particle behavior 
(diffusive 
for the Brownian and ballistic for the Newtonian dynamics) at short times and diffusive behavior at long times. Upon approaching 
the critical density $\wfd$, the diffusion starts slowing down until the diffusion coefficient vanishes at $\wfd$. A finite plateau,
$\lim_{t\to\io} \D(t) = \D_\io$, is observed in the MSD at long times for $\wh\f>\wfd$.
The long-time limit $\D_\io$ 
can be compared to the asymptotic result from the plateau equation, which is given by 
$\D_\io = T^2/\MM_\io$ from Eq.~\eqref{eq:MSD}~\cite{PUZ20}.

We next discuss the critical behavior of the diffusivity upon approaching the 
dynamical transition. It is expected from the asymptotic 
analysis of the 
DMFT equations~\cite{KPUZ13,PUZ20}
that the diffusion coefficient $\wh D$ defined in Eq.~\eqref{eq:D} follows 
a power-law behavior, \ie $\wh D \sim \vert \wfd - \wh\f \vert^\g$ when 
$\wh \f \to \wfd^-$, as in
Mode Coupling Theory (MCT)~\cite{Go09}. 
The critical exponent 
$\g$ can be computed for the SQS potential at $\wh\ee=10$, returning 
the value $\g \simeq 2.25844$~\cite{KPUZ13,PUZ20}. The critical behavior is clearly seen in Fig.~\ref{fig:diffusivity-SS}a. 
Interestingly, we find that 
in the Brownian case the power-law behavior is observed also for packing 
fractions rather distant from the critical point.
On the other hand, the Newtonian diffusivity is diverging for low densities
because of the ballistic motion in the absence of interactions. However, in 
both cases the agreement with the predicted critical scaling is very good; the difference in 
the prefactor is due to the difference in microscopic time units in Brownian and Newtonian dynamics.
Note that for the points closest to $\wfd$, the numerical estimate of $\MM(t)$ at long times is less precise, which explains the slight deviation
of the diffusivity
from the expected power-law.

In the $d\to\io$ limit, the diffusion coefficient and the viscosity are related by a generalized Stokes-Einstein relation~\cite{MKZ15,CCS18,PUZ20}.
Indeed, the rescaled shear 
viscosity defined in Eq.~\eqref{eq:etas} is given by the interaction
term only when $d \to \io$; therefore, Eqs.~\eqref{eq:D} 
and~\eqref{eq:etas} lead to a slightly modified Stokes-Einstein relation (SER)
\beq\label{eq:SER}
\b \wh D \wh\h_s = \frac{\b \wh\f \int_0^\io \de t \, \MM(t)}{\wh\z + \b \int_0^\io \de t \, \MM(t)} 
= \frac{\wh\f}{1+\wh\z \wh\f/\wh\h_s} \ ,
\eeq
which is plotted as a function of $\wh\f$ in Fig.~\ref{fig:diffusivity-SS}b.
While this relation is trivial for the Newtonian case, in which $\b \wh D \wh\h_s = \wh\f$ 
because $\wh\z=0$, the same is not true for Brownian dynamics, in which the 
linear behavior is only recovered asymptotically close to the dynamical transition, where the viscosity diverges.

\subsection{Brownian hard spheres}
\label{ssec:results-HSB}

\begin{figure}[t]
    \centering
    \includegraphics[width=0.47\textwidth]{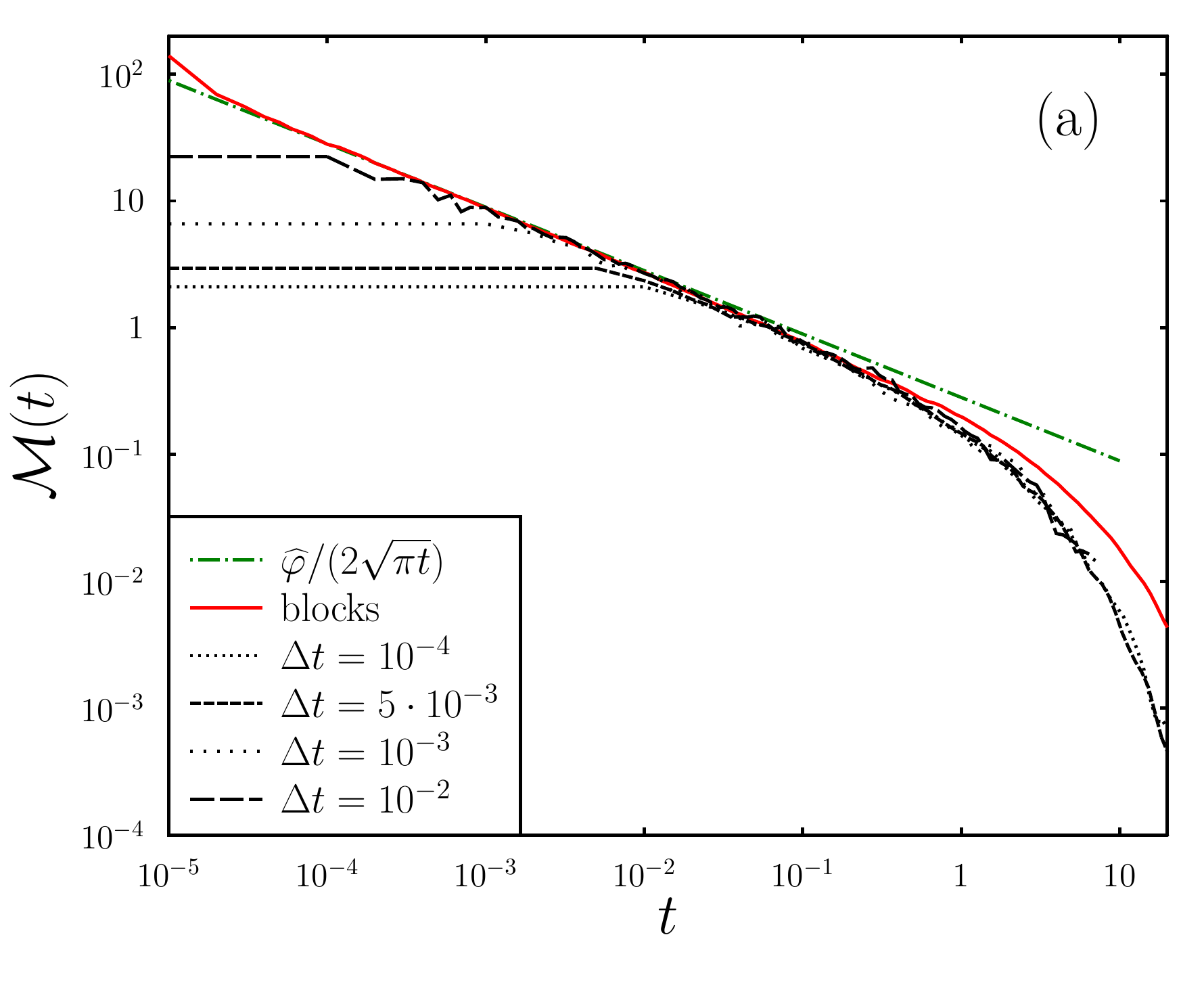}
    ~
    \includegraphics[width=0.47\textwidth]{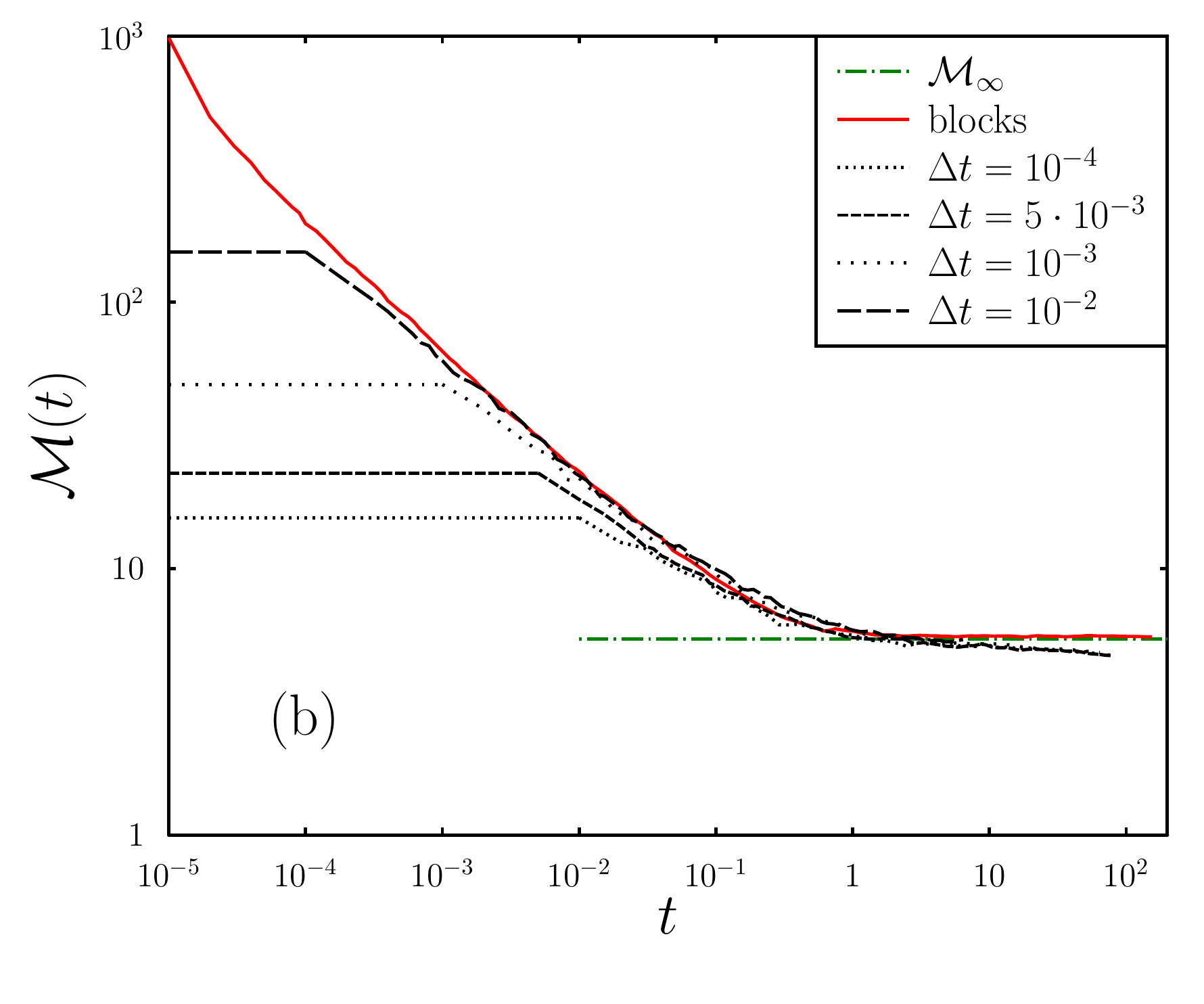}
    \caption{
    Numerical solution of the DMFT for Brownian hard spheres.
    Comparison between $\MM(t)$ obtained on a fixed grid (black lines) 
    and with a decimation algorithm (continuous red line), at $\wh\f=$1.0 (a) and 7.0 (b). The simulations with fixed time grid are plotted for several time steps $\D t$. The decimation algorithm has $\D t_{\min} = 10^{-5}$ and $\D t_\max = 2\cdot 10^{-2}$.
    The dot-dashed green lines represent the short-time behavior from 
    Eq.~\eqref{eq:firstHS} (left), and the analytical plateau $\MM_\io$ at $\wh\f=7.0$ (right) 
    respectively.}
    \label{fig:M-HSB}
\end{figure}

\begin{figure}[t]
    \centering
    \includegraphics[width=0.47\textwidth]{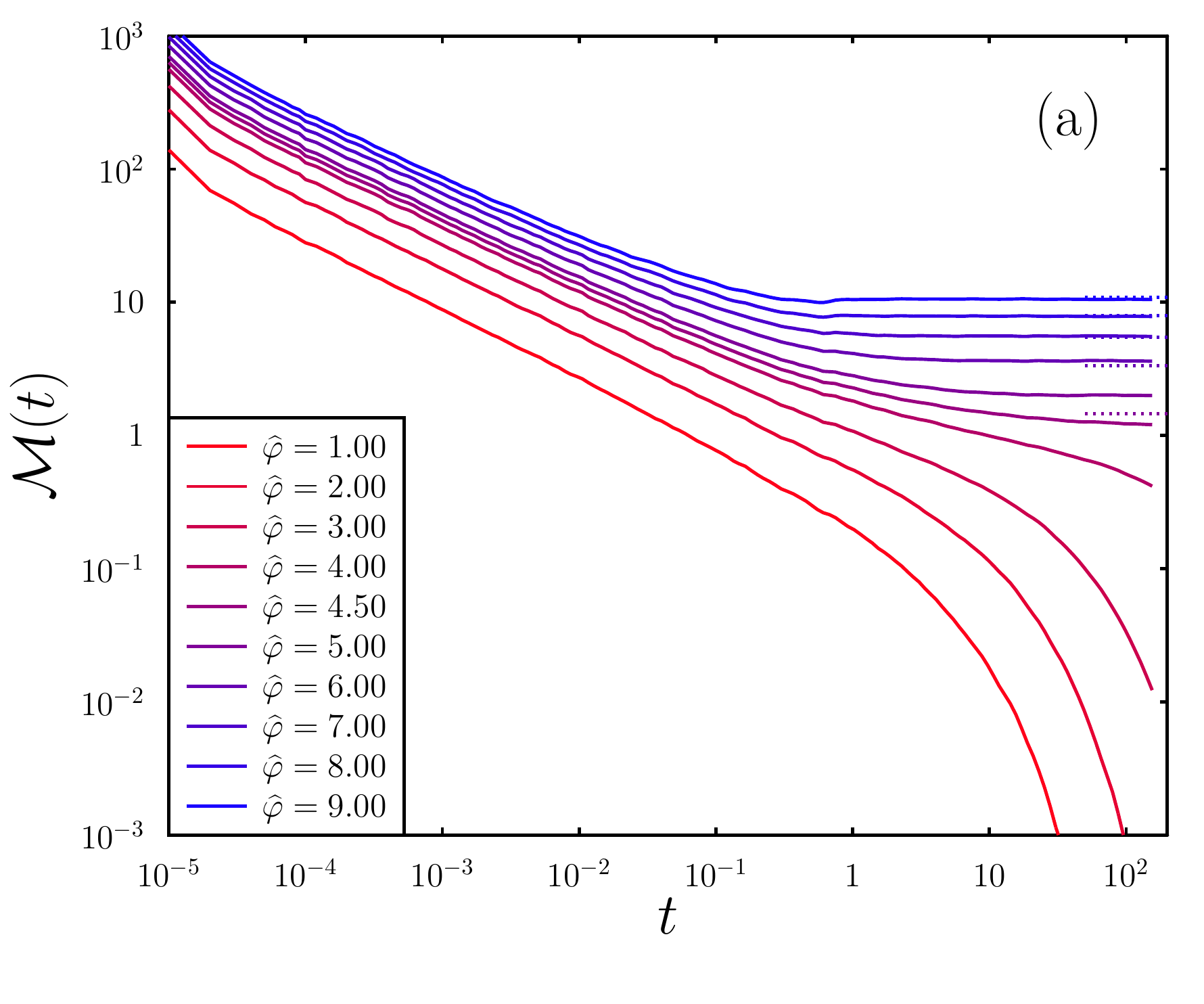}
    ~
    \includegraphics[width=0.47\textwidth]{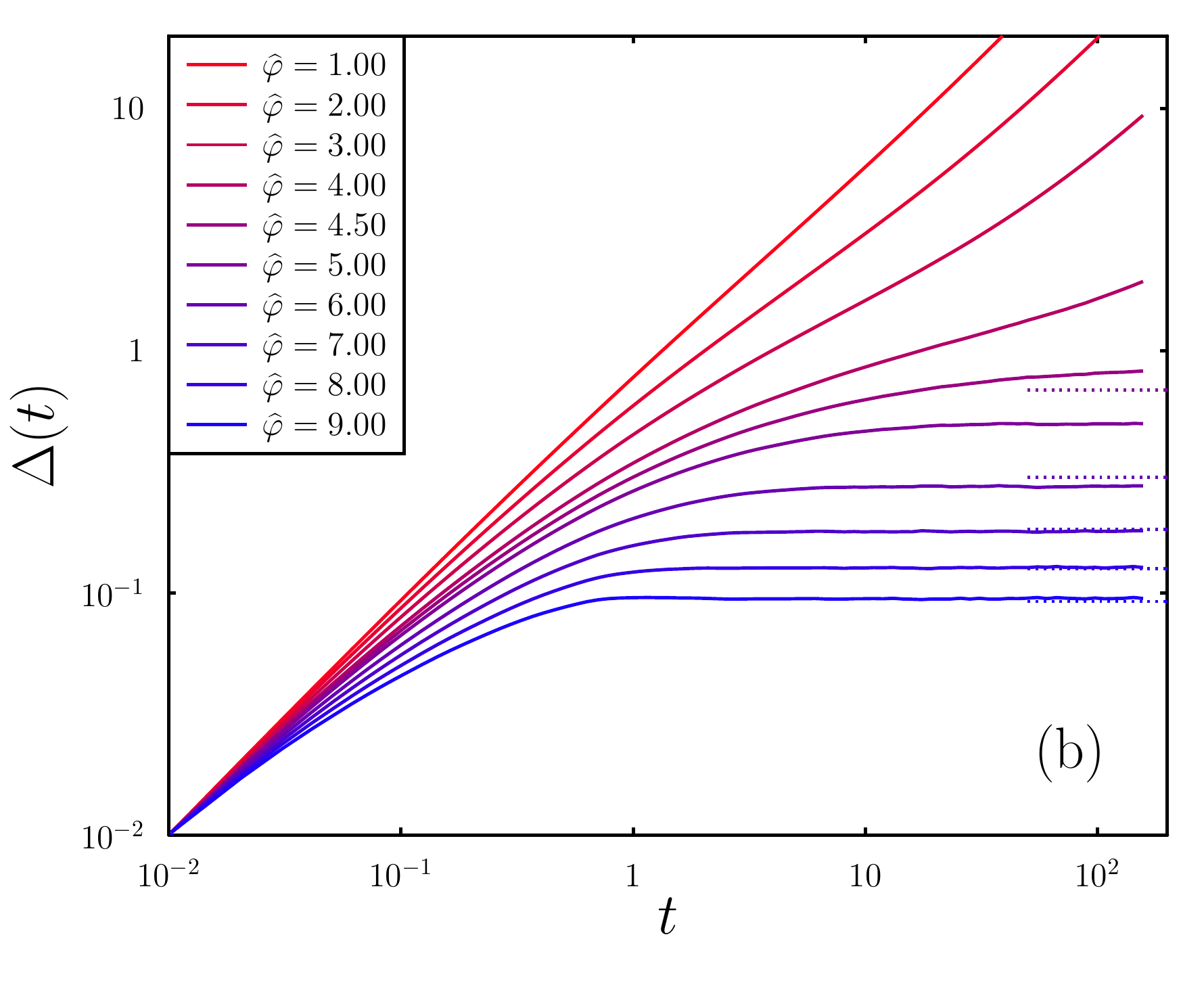}
    \caption{
    Numerical solution of the DMFT for Brownian hard spheres obtained via the decimation algorithm.
    (a) Memory function $\MM(t)$
    for several packing fractions given in the legend. Dashed lines correspond 
    to the plateaus $\MM_\io$ when $\wh\f > \wfd \simeq 
    4.8067$. (b) Mean square displacement $\D (t)$ computed
    from the corresponding $\MM(t)$, and comparison with the 
    plateau $\D_\io$.
    }
    \label{fig:HSB-blocks}
\end{figure}

The numerical results are less clear for Brownian hard spheres.
In this case, 
we know from Eq.~\eqref{eq:firstHS} that the short-time memory diverges as $t^{-1/2}$.
We find that 
the numerical solution for $\MM(t)$ depends on the discretization: in fact, the memory functions obtained through a fixed time grid and a decimation algorithm 
differ. The discrepancy is shown in Fig.~\ref{fig:M-HSB}, in which we show two cases with 
$\wh\f=1.0$ and $\wh\f=7.0$. Knowing that the critical density for hard spheres is 
$\wfd \simeq 4.8067$, we expect to observe a complete decay in the dilute case and a plateau 
in the dense case. While the fixed-grid solutions approach 
the decimation solution for short times, there is a clear discrepancy at long times, which does not 
seem to depend on the time step $\Delta t$ chosen for the fixed-grid algorithm. Moreover, the fixed-grid solution at $\wh\f=7.0$ slowly decays below the plateau, 
while the decimation algorithm solution is going to the expected plateau 
$\MM_\io \simeq 5.45227$ obtained from Eq.~\eqref{eq:MplateauHS}.
While we do not have a clear explanation for this discrepancy, we suspect that the short-time cutoff to the square root divergence imposed by the fixed time step affects the memory function even at long times. The decimation algorithm is able to partially cure this problem because the short-time part of the memory function is integrated more accurately.
The results obtained with the decimation algorithm are thus
closer to the expected asymptotic limits. 

The memory function $\MM(t)$ obtained via the decimation algorithm is shown in 
Fig.~\ref{fig:HSB-blocks}, for $\wh\f=1.0,2.0,\ldots,9.0$, and we observe 
a decay to zero in the liquid phase $\wh\f < \wfd$ and a plateau in the solid phase $\wh\f > \wfd$.
However, the results display two main issues: first, there is a clear jump in the solution 
around $t \sim 1$ when we rescale $\D t \to 2 \D t$, yielding an unphysical discontinuity in 
$\MM(t)$. Second, the plateau is far from that obtained from Eq.~\eqref{eq:MplateauHS} when $\wh\f \to \wfd^+$. 
The reason for these discrepancies is unclear, and we unfortunately must conclude that our numerical integration schemes are not reliable for Brownian hard spheres.

\subsection{Newtonian hard spheres}
\label{ssec:results-HSN}

\begin{figure}[t]
	\centering
	\includegraphics[width=0.47\textwidth]{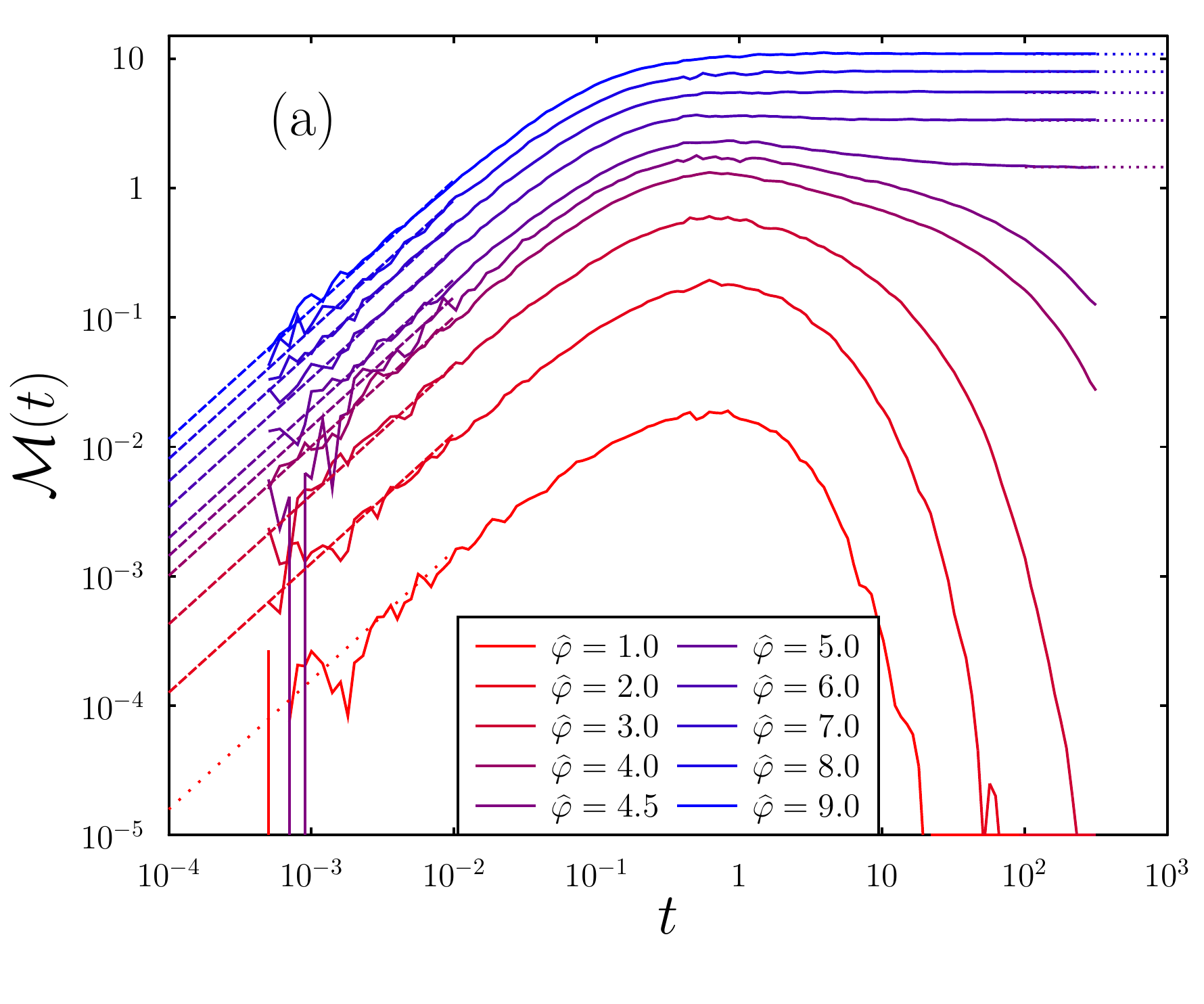}
	~
	\includegraphics[width=0.47\textwidth]{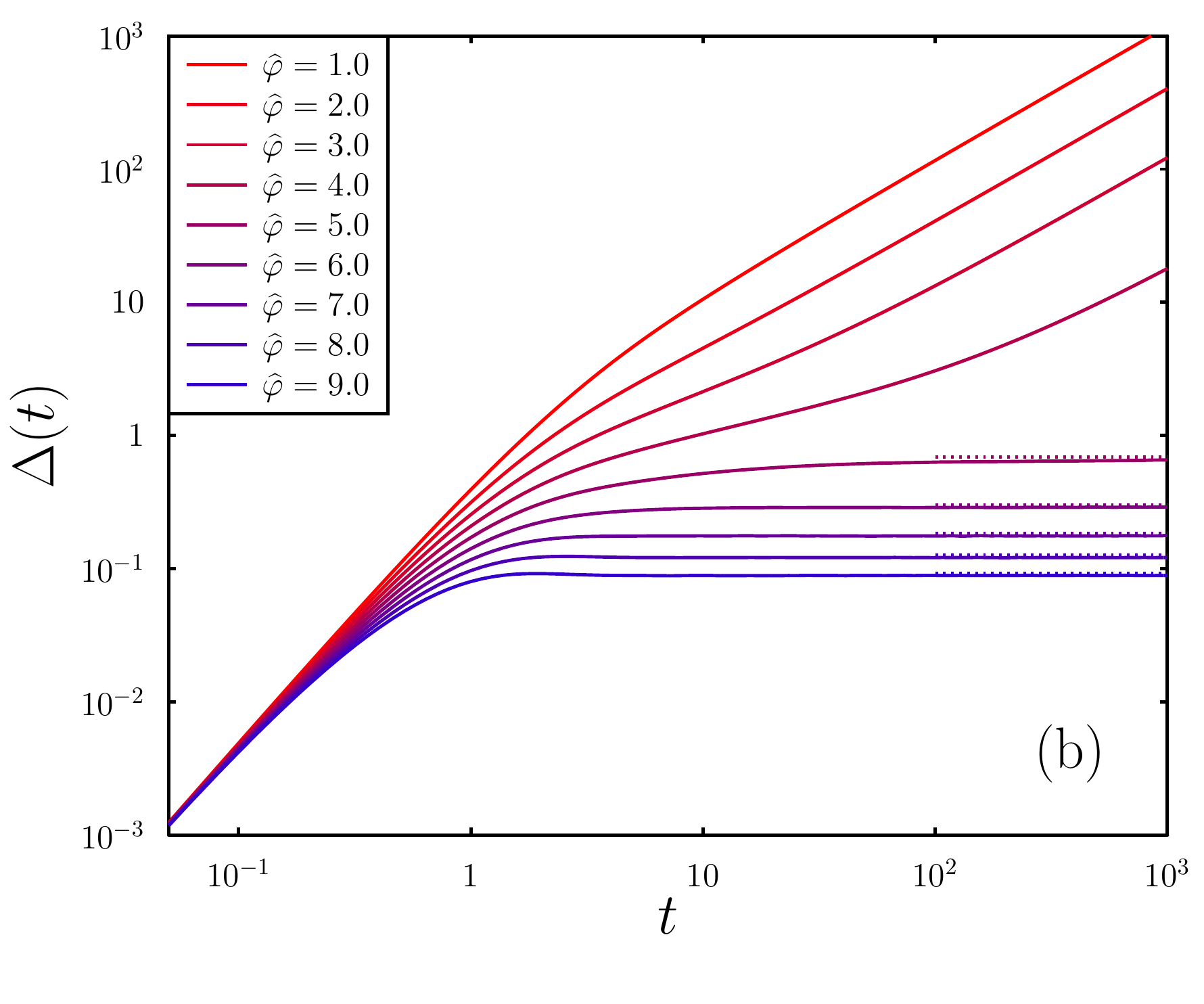}
	\caption{
	Numerical solution of the DMFT equations for Newtonian hard spheres. 
	(a)~Memory kernel $\MM(t)$.
	The numerical solution (continuous lines) 
	is compared to the short-time prediction $\MM(t) \sim C t$ with the constant $C$ obtained from Eq.~\eqref{eq:MregNewt_short} (dashed lines) and the 
	expected plateau in the dynamically arrested phase (dotted lines). The curves are obtained by 
	piecewise concatenation of the solution found with $\D t = 10^{-4}$ up 
	to $t_\max =1$ (short times) and the solution with $\D t= 5\cdot 
	10^{-3}$ and $t_\max \sim 3\cdot 10^{2}$.
	(b)~Mean square displacement computed from the numerical 
	$\MM(t)$. The dotted lines correspond to the static plateau $\D_\io = 
	T^2/\MM_\io$.}
	\label{fig:HSN}
\end{figure}

\begin{figure}[ht!]
	\centering
	\includegraphics[width=0.47\textwidth]{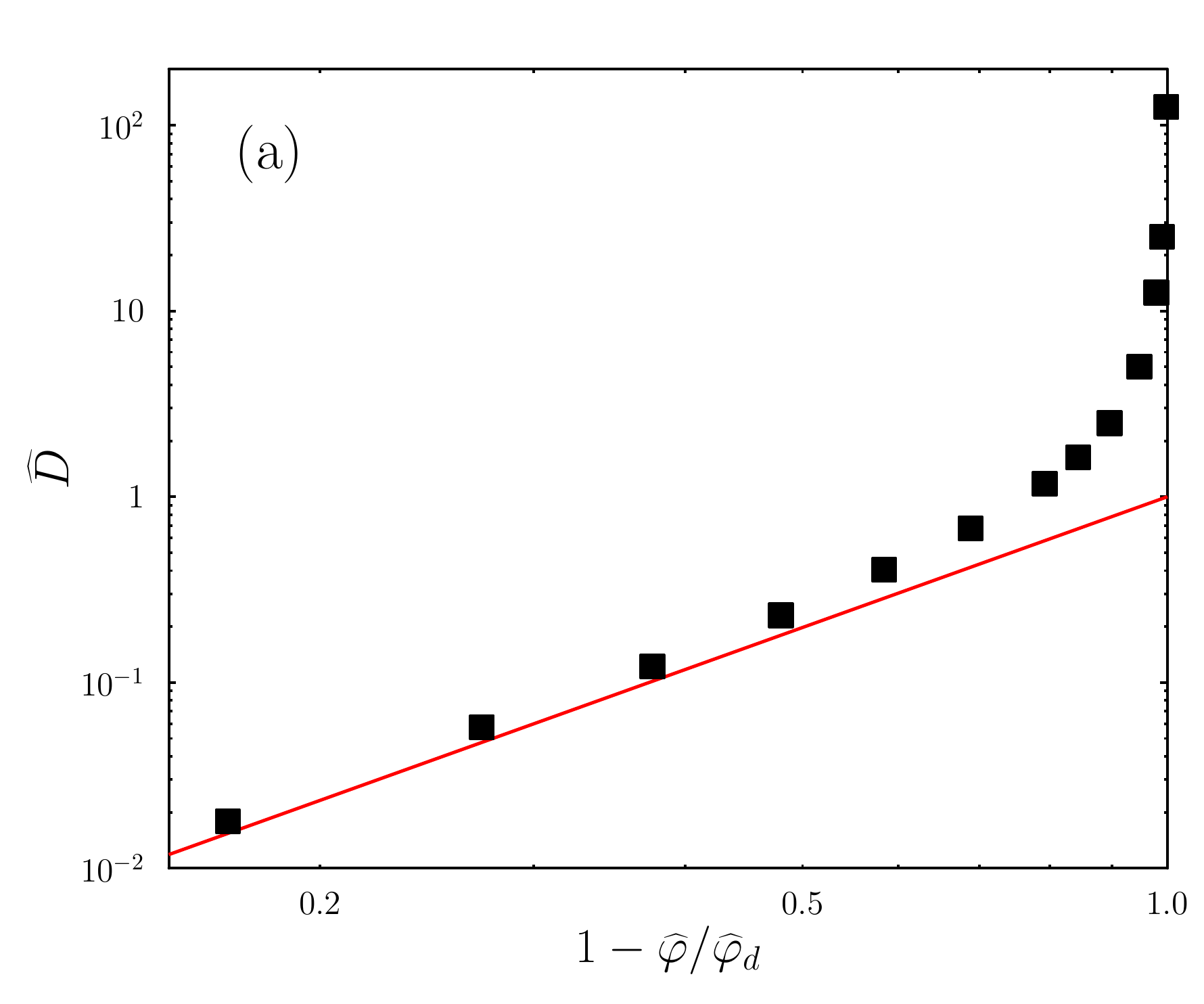}
	~
	\includegraphics[width=0.47\textwidth]{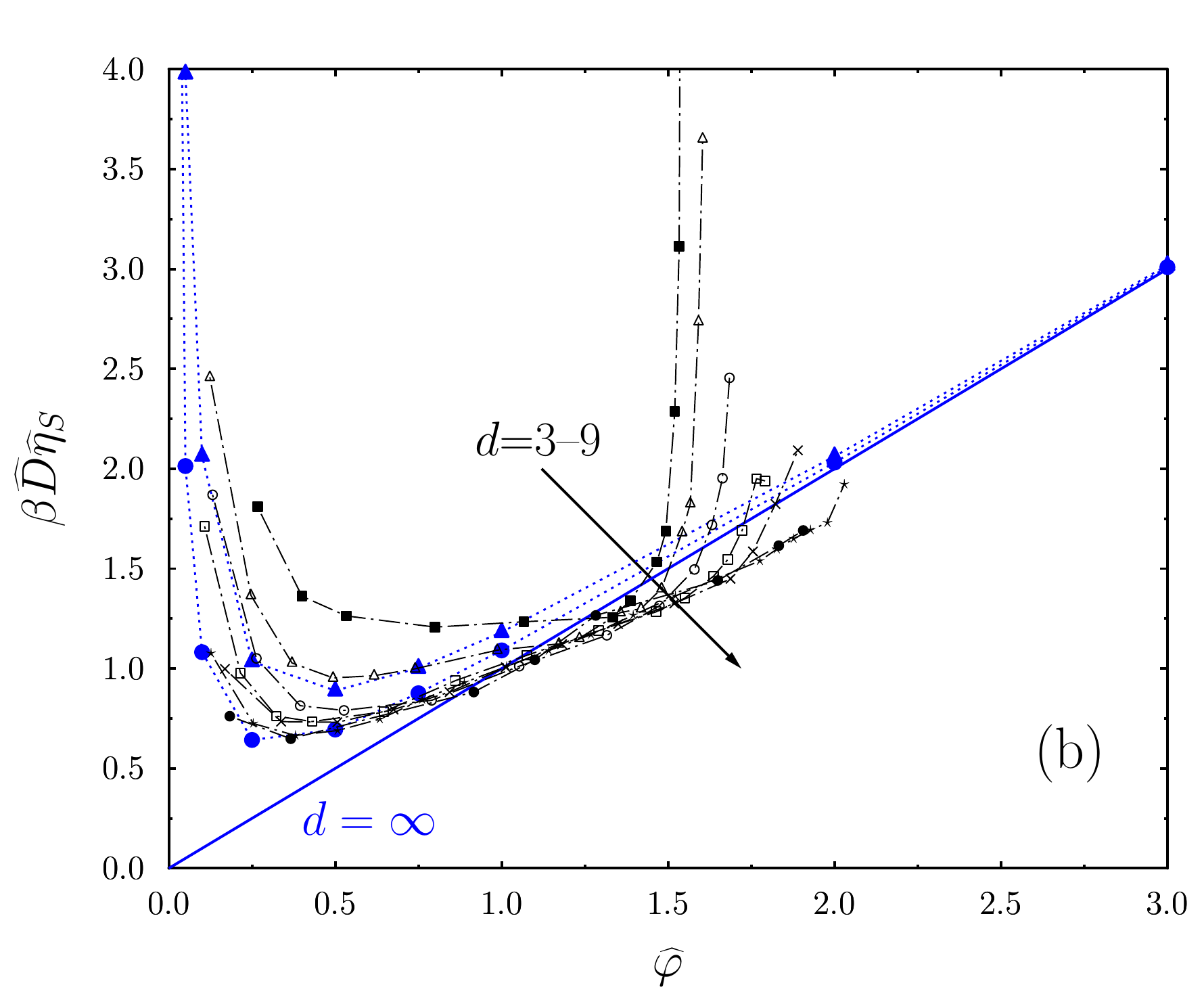}
	\caption{
	Numerical solution of the DMFT equations for Newtonian hard spheres. 
	(a)~Critical scaling of the diffusivity $\wh D \sim (\wfd - \wh\f)^\g$
	    for $\wh\f \to \wfd^-$, as in Fig.~\ref{fig:diffusivity-SS}a but with $\wfd=4.8067$ and $\g = 2.33786$.
	    (b)~Stokes-Einstein relation including the kinetic viscosity 
	    for Newtonian hard spheres, computed from 
	    Eq.~\eqref{eq:SER-num} with $d=8$ and $d=12$ (blue points) and its
	    $d=\io$ limit (blue line). The black points correspond to MD 
	    simulations with $3\leq d \leq 9$, from~\cite{CCJPZ13}.
	 }
	\label{fig:diffusivity-HSN}
\end{figure}

The fixed time grid algorithm works well 
when considering Newtonian hard spheres. Indeed, in this 
case the memory kernel can be separated into a singular and a regular part, see 
Eq.~\eqref{eq:HSNseparation}. The singular part provides a white noise contribution
in the dynamics in Eq.~\eqref{eq:MregNewt2} which can be discretized in a standard way, 
so we only need to compute self-consistently the regular part $\MM_{\rm reg}(t)$. We recall that a short-time exact analysis gives $\MM_{\rm reg}(t) \sim 0.1578 \, \wh\f^3 \, t$,
see Appendix~\ref{app:NewtonianHS2}. At long times,
the memory function is expected to exhibit a plateau for $\wh\f > \wfd \simeq 4.8067$. These asymptotic results are 
well reproduced by the numerical solution, as it can be seen in Fig.~\ref{fig:HSN}a. The mean square displacements is shown in
Fig.~\ref{fig:HSN}b. 
The scaling of the diffusivity is shown in Fig.~\ref{fig:diffusivity-HSN}a.
The critical scaling $\wh D \sim |\wh\f_d - \wh\f |^\g$ is confirmed, with the expected critical exponent $\g \simeq 2.33786$~\cite{KPUZ13}.

In this paper we did not report any direct comparison between the $d\to\io$ results and numerical data in finite~$d$ because the latter are only available for $d\leq 10$ and finite $d$ corrections are usually too large to prevent a quantitative comparison with the $d\to\io$ solution, except for the critical scaling around the dynamical glass transition~\cite{CCPZ12,CCJPZ13}.
However, an exception is given by
the Stokes-Einstein relation for Newtonian dynamics, for which the finite $d$ corrections seem unusually mild~\cite{CCJPZ13}, which motivated us to attempt a systematic comparison with the solution of the DMFT equations. 
Note that in $d\to\io$ the exact solution trivially gives
$\b \wh D\wh \h_s = \wh\f$ for any potential. 
However, the kinetic term provides a $1/d$ correction that diverges in the dilute regime, leading to
\beq\label{eq:SER-num}
\b \wh D \wh \h_s = \wh\f \left[ 1 + \frac{1}{d} \frac{\int^\io_0 \de u \, \ddot \D(u) ^2}{\int^\io_0 \de u \, \MM(u)} \right] \ ,
\eeq
where the integrals are computed on the dimensionless times $u = t/\t_N$ and memory functions 
$\MM(u) = \b^2 \MM(u \cdot \t_N)$.
The kinetic term vanishes in the $d \to \io$ limit, but it also diverges in the dilute limit when $\wh\f\to 0$ because the motion becomes ballistic at all times and the integral of $\ddot \D(t)^2$ diverges (see Appendix~\ref{app:NewtonianHS3}).
In Fig.~\ref{fig:diffusivity-HSN}b 
we plot our results from Eq.~\eqref{eq:SER-num} together with finite-dimensional data from Ref.~\cite{CCJPZ13}.
The divergence of
$\b \wh D \wh \h_s$ in the low-density regime is well visible, and the region over which it is observed shrinks upon increasing $d$, as expected. At finite value of $\wh\f$, the results from Eq.~\eqref{eq:SER-num} thus converge to the line $\b \wh D\wh \h_s = \wh\f$ when $d\to\io$.
The finite-$d$ numerical data agree well with Eq.~\eqref{eq:SER-num} in the low-density regime, while they seem to accumulate on a straight line with slope slightly smaller than one at higher density. We attribute this difference to the fact that $d$ is not large enough, so other subleading terms in $1/d$ could play a role. 
Note that the value of $\wfd$ for $d=8$ is around $\wfd \sim 2$, hence still quite distant from the asymptotic limit. 
In the vicinity of $\wfd$, a steep increase of $\b \wh D\wh \h_s$ above the mean field prediction is observed. This ``Stokes-Einstein relation breakdown" 
is a genuinely non-mean-field effect, related to dynamical heterogeneities~\cite{BB07,BBBCS11}, and it vanishes in the limit $d\to\io$~\cite{CCJPZ13}.

\section{Conclusions}
\label{sec:conclusions}

In this work, we analyzed the DMFT equations for infinite-dimensional equilibrium liquids derived in~\cite{MKZ15,Sz17,AMZ18}.
We derived the DMFT equations for hard spheres as a limit of those for regular potentials, and
we presented some methods to solve the DMFT equations numerically and analyze them analytically in some asymptotic limits. 
Our numerical solution algorithm is based on a straightforward discretization of time and an iterative calculation of the kernel $\MM(t)$ via the self-consistent condition.

For soft spheres (both Brownian and Newtonian) and Newtonian hard spheres, we obtain accurate numerical solutions 
which agree well with the expected asymptotic limits (short times, long times, low density). 
The results confirm the presence of a dynamical glass transition with the same critical properties as Mode-Coupling Theory (although with different exponents), and provide the shape of the memory function and of the mean square displacement both in the liquid and glass phases. For Brownian hard spheres, the numerical integration scheme seems unable to properly handle the short-time divergence of the memory function, and the resulting numerical solutions are not fully consistent with the asymptotic limit. Better discretization schemes should then be developed, which is a non-trivial problem in stochastic calculus.

Unfortunately, the algorithm is limited to relatively short times, as it is often the case in the study of DMFT equations, which prevents us to investigate long-time phenomena such as the dynamic criticality around the glass transition (e.g. the stretching exponent of the memory function) and the aging dynamics in the glass phase~\cite{CK93,Cu02,FFR19}.
A possible improvement would be the implementation of a decimation 
algorithm, which allows one to take exponentially growing time steps in the numerics 
and to observe the dynamics over several decades of time. Altough this 
method is well-established, \eg in the numerical solution of Mode-Coupling 
Theory equations~\cite{FGHL91jcp,LOV17pre,GPF20pre}, its extension 
to stochastic dynamics is missing. Such algorithm would represent a powerful tool for future 
investigations.

This work opens the way to the numerical solution of the DMFT equations in the non-equilibrium case~\cite{AMZ18,AMZ19}, which hopefully will give insight on a variety of phenomena such as yielding, jamming, and glass melting, both in passive and active systems.
The simplest case is that of active matter in infinite dimensions~\cite{PLW19}. While the approach of Ref.~\cite{PLW19} focuses on the stationary state distribution, 
DMFT  also describes time-dependent correlations, as in the MCT approach of Refs.~\cite{BK13,SFB15}, and the approach to the stationary state itself. 
Work is currently in progress to solve, both analytically and numerically, the DMFT equations for the same model investigated in Ref.~\cite{PLW19}.
Another interesting case is that of rheology. Recently, MCT has been extended to describe the rheology of liquids and glasses, 
either in stationary state in the schematic limit~\cite{BBK00}, or 
via an integration-through-transient approach in the general setting~\cite{FC02,BVFLC09,BCF12}. The same problem can be approached more phenomenologically
via elastoplastic models~\cite{NFMB17}. It would be very interesting to compare DMFT with these complementary approaches.

Finally, a very important direction for future research is that of understanding the finite-$d$ corrections to DMFT in a systematic way. 
Ref.~\cite{BR19} reported a numerical calculation of $\MM(t)$ in $d=3$, and a direct comparison with its DMFT approximation. This study should provide
important insight on which terms have to be added to DMFT to obtain a more quantitative theory in finite $d$~\cite{JR15,CCS18}. Several groups are working in
this direction to formulate a ``cluster DMFT''~\cite{KSPB01} of the glass transition.
Another independent direction to tackle the same problem is that of looking for systematic deviations between DMFT and numerical results in high dimensions~\cite{BCCHHSZ20}.
This should allow one to identify the dominant corrections, such as hopping events (instantonic corrections)~\cite{CJPZ14}, facilitation effects~\cite{BBW08},
or fluctuations due to disorder~\cite{BB07,SBBB09,FPRR11,RV15,RV19}.

\subsection*{Data availability statement}
The data that support the findings of this study are available from the corresponding author upon reasonable request.

\acknowledgments

We warmly thank P.~Charbonneau, M.~Fuchs and T.~Franosch for many useful exchanges, E.~Agoritsas, G.~Biroli, J.~Kurchan,
T.~Maimbourg, G.~Szamel and P.~Urbani for many discussions about the theoretical modeling, 
and F.~Roy for many insights about the numerical solution of the DMFT equations. 
We thank the referees for providing many useful suggestions that improved considerably the paper after the first revision.
This project has received funding from the European Research Council (ERC) under the European Union's Horizon 2020 research and innovation programme (grant agreement n. 723955 - GlassUniversality).

\appendix
\section{Some details on Newtonian dynamics}

We provide here some additional details on the formulation of the DMFT equations in the Newtonian case.

\subsection{Distribution of the initial velocity}
\label{app:Maxwell}

The distribution of the initial velocity $\dot h_0$ was not specified in~\cite{MKZ15,AMZ18,AMZ19} because it easily follows from the Maxwellian statistics of velocities in equilibrium. We provide
some details here for completeness.
Following~\cite[section 5.1]{AMZ18}, 
we define $y(t) = (d/\ell) \hat{\mathbf{r}}_0 \cdot (\mathbf{u}_1(t)-\mathbf{u}_2(t))$, where
$\hat{\mathbf{r}}_0$ is the unit vector along the initial distance between two particles (essentially a random unit vector by isotropy),
and $\mathbf{u}_{1,2}$ are the displacements of the two particles with respect to their initial position at time $t=0$.
According to the Maxwell distribution, each component of $\dot{\mathbf{u}}_{1,2}$ is an independent Gaussian variable with zero mean and variance $T/m$. As a consequence, $\dot y(t)$ is also a random Gaussian variable with zero mean and variance $(d^2/\ell^2) 2 T/m = T/\wh m$. 
Finally, $h(t) = h_0 + y(t) + \D(t)$ and $\D(t)$ is ballistic at short times, hence $\dot \D(0)=0$, which implies that $\dot h_0 = \dot h(0) = \dot y(0)$ is also a Gaussian variable with zero mean and standard deviation~$T/\wh m$. This justifies the probability distribution of $\dot h_0$ in Eq.~\eqref{eq:Mdef}.

\subsection{Viscosity}
\label{app:etas}

The derivation of the viscosity in $d\to\io$ is discussed in~\cite{MKZ15,PUZ20} where, however, the kinetic term has been omitted. We provide here a more detailed discussion that is needed to compare with simulation results.

The shear viscosity is given in terms of the autocorrelation of the stress tensor in~\cite[Eq.(8.4.10)]{Hansen}. The stress tensor, as given in~\cite[Eq.(8.4.14)]{Hansen}, is the sum of a kinetic and an interaction terms. The contribution of the autocorrelation of the interaction term has been discussed in~\cite[section 3.4.1]{PUZ20}, and is given by
$\h_s = \b \r d \int_0^\io \de t \, \MM(t)$.
The other terms are subdominant when $d\to\io$, but the autocorrelation of the kinetic term provides a divergent contribution for $\wh\f\to0$ that we need to add if we want to properly reproduce the ideal gas limit.

We are then going to neglect the cross-correlation of the kinetic and interaction terms, because it is subdominant both in $1/d \to 0$ and in $\wh\f\to 0$, as deduced from ~\cite[Eq.(8.4.21)]{Hansen} and~\cite[Eq.(D2)]{CCJPZ13}.
The autocorrelation of the kinetic term can be written, neglecting velocity correlations between distinct particles in the limit $\wh\f\to0$, as
\beq\label{eq:ZvsD}
\h_{\rm K} = \b\r m^2 \int_0^\io \de t \, Z(t)^2 \ ,
\qquad Z(t) = \la v_\m(t) v_\m(0) \ra
= \frac1{2d}\frac{\de^2}{\de t^2} 
\la | \mathbf{r}(t) - \mathbf{r}(0) |^2 \ra
\ ,
\eeq
where $Z(t)$ is the autocorrelation of a single spatial component of the velocity~\cite[Eq.(7.2.1)]{Hansen}, related to the mean square displacement
by~\cite[Eq.(7.2.5)]{Hansen} (here we considered a representative particle, without indicating explicitly the average over the $N$ particles).
Recalling the scaling of mass and mean square displacement defined in section~\ref{sec:DMFE}, we obtain
\beq
\h_{\rm K}=\b \r \wh m^2 \int_0^\io \de t\, \ddot \D(t)^2 \ .
\eeq
Summing the kinetic and interaction contributions, we obtain Eq.~\eqref{eq:etas}.

\section{Brownian hard spheres}

\subsection{Backward Kolmogorov equation after the first iteration}\label{app:BrownianHSBack}

We first provide some details on the backward Kolmogorov equation satisfied by the central quantity $G(h_0,t) =  \langle \theta(-h)\rangle_{h_0} $, i.e. the probability for a particle starting from $h_0$ to end up on the negative axis at time $t$. Here we consider the first iteration for the Brownian case, as studied in section~\ref{ssec:HSB}. By definition, this probability $G(h_0,t) = \langle \theta(-h)\rangle_{h_0} $ can be computed by integrating the propagator $p_{\wh\ee}(h, t | h_0)$ over the final position, i.e.,
\beq\label{def_G_app}
G(h_0,t) = \langle \theta(-h)\rangle_{h_0} = \int_{-\infty}^0 \de h \, p_{\wh\ee}(h, t | h_0) \;.
\eeq
A natural way to compute $G(h_0,t)$ would then be to write the standard (i.e. {\it forward}) Kolmogorov equation for the propagator $p_{\wh\ee}(h, t | h_0)$, which, roughly speaking, amounts to study the dependence of $p_{\wh\ee}(h, t | h_0)$ on the {\it final position} $h$. One would then solve this forward equation to obtain the propagator $p_{\wh\ee}(h, t | h_0)$, insert this expression in (\ref{def_G_app}) and integrate over the final position $h \in (-\infty,0)$ to get finally $G(h_0,t)$. In such cases, there exists however a simpler way to do this computation, which avoids the explicit evaluation of the integral over the final position. It amounts to write instead the {\it backward} Kolmogorov equation for $p_{\wh\ee}(h, t | h_0)$, which describes the dependence of $p_{\wh\ee}(h, t | h_0)$ in 
the {\it initial} position $h_0$ (see \eg\cite{BMS13}). For a general Langevin equation as in \eqref{eq:Brow_drift} of the form 
\begin{equation}\label{eq:Brow_drift_gen}
\dot h(t) 
= F[h(t)] + \X(t)  \ , \qquad   \langle \X(t) \X(u) \rangle  = 2 \, \d(t-u) \ ,
\end{equation}
for some force field $F[h(t)]$, it is well known that this backward Kolmogorov equation reads
\begin{equation}\label{back_Kolmo_p}
\partial_t p_{\wh\ee}(h, t | h_0) = \frac{\partial^2}{\partial h_0^2} p_{\wh\ee}(h, t | h_0) + F[h_0] \frac{\partial}{\partial h_0} p_{\wh\ee}(h, t | h_0) \;.
\end{equation}
The interesting feature of this backward approach is that, by integrating this equation \eqref{back_Kolmo_p} over the final position $h \in (-\infty,0)$, one finds that $G(h_0,t)$ in \eqref{def_G_app} actually satisfies exactly the same equation \eqref{back_Kolmo_p}, namely
\begin{equation}\label{back_Kolmo_G}
\partial_t G(h_0,t) = \frac{\partial^2}{\partial h_0^2} G(h_0,t) + F[h_0] \frac{\partial}{\partial h_0} G(h_0,t) \;.
\end{equation}
By specifying this equation \eqref{back_Kolmo_G} to the case $F[h_0] = 1 + \hat \varepsilon \theta(-h_0)$, one obtains the equations given in \eqref{eq:Gev}. The initial and boundary conditions in \eqref{eq:Aboundary} are then obtained by natural physical considerations.   

\subsection{Derivation of the Laplace transform of the first iteration}
\label{app:BrownianHS}

We provide here some details on the solution of Eq.~\eqref{eq:Gev} with boundary conditions in Eq.~\eqref{eq:Aboundary}.
Consider the Laplace transform $\wt G(h_0,s) = \int_0^\io \de t e^{-st} G(h_0,t)$.
Using an integration by parts,
\beq
\int_0^\io \de t e^{-st} \dot G(h_0,t) = s \wt G(h_0,s) + [e^{-s t} G(h_0,t)]_0^\io = s \wt G(h_0,s) - G(h_0,0) \ ,
\eeq
and taking into account the first boundary condition in Eq.~\eqref{eq:Aboundary}, we get
\beq\label{eq:Laplace}
\begin{split}
s \wt G(h_0,s) -1 &=  \wt G''(h_0,s) +(1+\wee) \wt G'(h_0,s) \ , \qquad h_0 < 0 \ , \\
s \wt G(h_0,s) &= \wt G''(h_0,s) + \wt G'(h_0,s) \ ,\qquad \qquad\quad h_0 > 0 \ .
\end{split}\eeq
Taking into account the two other boundary conditions in Eq.~\eqref{eq:Aboundary}, the solution is
given by Eq.~\eqref{eq:GLap} with yet unknown functions $c_\pm(s)$.
Now we should impose the continuity conditions in $h_0=0$. The leading singularity in Eq.~\eqref{eq:Laplace} is of the form
$\wt G''(h_0,s) \approx \th(h_0)$ (note that a jump singularity comes from both the right and left hand sides of the equation), which implies
that $\wt G(h_0,s)$ and $\wt G'(h_0,s)$ are both continuous functions of $h_0$. This gives the conditions
\beq\begin{split}
\frac1s + c_-(s) &= c_+(s) \ , \\
c_-(s) \l_-(s)  &= c_+(s) \l_+(s)  \ ,
\end{split}\eeq
which imply
\beq\label{eqA:cpm}
c_-(s) = \frac{\l_+(s)}{s [\l_-(s) - \l_+(s)]} \ ,
\qquad
c_+(s) = \frac{\l_-(s)}{s [\l_-(s) - \l_+(s)]} \ ,
\eeq
thus completing the proof of Eq.~\eqref{eq:GLap}.

\subsection{Short-time behavior of the stress-stress correlation}
\label{app:MHSshort}

The short-time behavior of the stress autocorrelation function $C_{\s\s}(t) = \b \la \Pi_{\m\n}(t) \Pi_{\m\n}(0) \ra/V$~\cite{Hansen,PUZ20} for Brownian hard spheres is given by kinetic theory for $d=3$~\cite{LR94,VDFC97,LCPF09}, as
\beq\label{eq:Cshort1}
C_{\s\s}(t\to 0) = \frac{18}5 \f^2 g(\ell) \frac{\h_0}{\t} \sqrt{\frac{2\t}{\pi t}}
= \frac{24}{5\pi} \f^2 g(\ell) \frac{T}{\ell^3} \sqrt{\frac{2\t}{\pi t}}
\ ,
\eeq
where $\ell$ is the sphere diameter, $\f$ is the packing fraction, $g(\ell)$ is the contact value of the pair correlation function, $D_0 = T/\z$ is the free particle diffusion coefficient, $\t = \ell^2/(4 D_0)$, and
$\h_0$ is given by the Stokes expression $D_0 = T/(3\pi \h_0 \ell)$.

According to~\cite{MKZ15,PUZ20}, when $d\to \io$, 
the stress-stress autocorrelation is simply related to the memory function by
\beq\label{eq:Cshort2}
C_{\s\s}(t) = \b \r d \MM(t)  \ , \qquad \MM(t\to 0) \sim \frac{\wh\f}2 T^2 \sqrt{\frac{\t_B}{\pi t}} 
\qquad\Rightarrow\qquad C_{\s\s}(t\to 0) \sim \frac{ 2^{2d} \G(1+d/2) }{2 d \pi^{d/2}}\f^2 \frac{T}{\ell^d} \sqrt{\frac{2\t}{\pi t}}  \ ,
\eeq
where we used the short-time result for Brownian hard spheres reintroducing physical dimensions, with
$\t_B = \wh\z/T=(\ell^2/2 d^2)\z/T=(\ell^2/2 d^2)/D_0=2 \t/d^2$, and $\r = 2^d \f/(V_d \ell^d)$ with
$V_d = \pi^{d/2}/\G(1+d/2)$.

The two expressions in Eq.~\eqref{eq:Cshort1} and \eqref{eq:Cshort2} match if we recall that $g(\ell)\to 1$ when $d\to \io$~\cite{PUZ20}, and if we interpret the factor $5$ as $d+2$ in Eq.~\eqref{eq:Cshort1}. This leads us to conjecture that in generic dimension $d$,
\beq\label{eq:B7}
C_{\s\s}(t\to 0) \sim \frac{ 2^{2d} \G(1+d/2) }{2 (d+2) \pi^{d/2}}\f^2 g(\ell) \frac{T}{\ell^d} \sqrt{\frac{2\t}{\pi t}}  \ ,
\eeq
which coincides with Eq.~\eqref{eq:Cshort1} in $d=3$ and with Eq.~\eqref{eq:Cshort2} when $d\to\io$, hence $d+2\approx d$.
We were unable to find Eq.~\eqref{eq:B7} in the literature, but it should follow from a straightforward generalization of the results of Refs.~\cite{LR94,VDFC97,LCPF09}
to arbitrary $d$.

\subsection{Low-density limit of the diffusion coefficient}
\label{app:Dlowphi}

The calculation of the diffusion coefficient for Brownian hard spheres 
has been previously done for systems up to three dimensions~\cite{HHK82,AF82jcp,LD84jcp}. A first method consists in solving the 
Smoluchowski equation for the relative dynamics of two particles, and 
get the mean-square displacement and the diffusion coefficient from 
the calculation of the self-intermediate scattering function and the memory 
function; the second method requires to compute the mobility $\mu$ of a 
tagged particle under the action of a small force $\mathbf{F}$, and use the 
Einstein relation $D = T \mu$.
Both methods agree that, when $d=2,3$, one finds at the first order in $\varphi$
\beq
D = D_0 ( 1 - 2 \varphi) \ ,
\eeq
being $D_0$ the diffusion at $\varphi=0$ and $\varphi$ the packing 
fraction~\cite{LD84jcp}.
 
This second method can be extended to any dimension $d$: from~\cite[Eq.~(17)]{LD84jcp}, the pair distribution function $g(\rr)$ has the general 
form
\beq
g(\rr) = e^{-\b v(r)} \left[ 1 + \b \ell \frac{Q(r)}{4r} \rr \cdot \mathbf{F} \right] \ .
\eeq
The radial function $Q(r)$ is determined by 
the differential equation in~\cite[Eq.~(16)]{LD84jcp}, which can be generalised to arbitrary dimension $d$ as
\beq
\frac{\de}{\de r} \left( e^{-\b v(r)} r^{d-1} \frac{\de Q}{\de r} \right) - (d-1) 
r^{d-3} Q(r) e^{-\b v(r)} + 2 \frac{r^{d-1}}{\ell} \frac{\de}{\de r} e^{-\b v(r)} = 0 \ ,
\eeq
with the boundary conditions $Q(r=\io)=0$ and $Q'(r=\ell)=-2/\ell$ for a hard sphere potential, leading to the general solution 
\beq
Q(r) = \frac{2}{d-1} \left( \frac{\ell}{r} \right)^{d-1} \th(r-\ell) \ .
\eeq
The following result can be used to compute the force $\mathbf{F}_{\text{relax}}$ 
exerted on the tagged particle by the surrounding ones, given by~\cite[Eq.~(24)]{LD84jcp}, \ie
\beq
\mathbf{F}_{\text{relax}} = \rho \int \de \rr \, g(\rr) \, v'(r) \frac{\rr}{r} = 
- \frac{2^d}{2(d-1)} \varphi \, \mathbf{F} \ .
\eeq
The total force acting on the tagged particle is therefore 
$\mathbf{F}_{\text{total}} = \mathbf{F} + \mathbf{F}_{\text{relax}}= [1 - 2^d \varphi / (2(d-1))] \mathbf{F} $. The 
coefficient of $\mathbf{F}$ coincides with the low-density correction to the 
mobility $\m$ and the diffusion coefficient then reads
\beq
D = D_0 \left( 1 - \frac{2^d}{2(d-1)} \varphi\right) = 
\begin{cases}
	D_0 \left( 1 - 2\varphi \right) &d=2, \, 3  \\[.2cm]
	D_0 \left( 1 - \frac12 \widehat\varphi \right) &d \to \io 
\end{cases} \ ,
\eeq
recalling that $\widehat\varphi = 2^d \varphi /d$. This result generalizes 
the low-density correction obtained for $d=2,3$ and extends it towards 
the infinite-dimensional limit, 
consistently with the DMFT result given in Eq.~\eqref{eq:D1_HSB}.

\section{Newtonian hard spheres}

\subsection{Short-time expansion from kinetic theory}
\label{app:NewtonianHS1}

In infinite dimensions, the memory function is related to the velocity autocorrelation~\cite{BR19}. 
We follow the notations of~\cite{PUZ20} and denote the non-scaled mean square displacement by $\DE(t) = \la | \mathbf{r}(t) - \mathbf{r}(0) |^2 \ra$. According to Eq.~\eqref{eq:ZvsD}, the velocity autocorrelation is $Z(t) = \ddot \DE(t)/(2d)$~\cite{Hansen}.
In the infinite dimensional limit, for Newtonian dynamics,
the non-scaled memory function $M(t) = (2d^2/\ell^2)\MM(t)$ is related to $\DE(t)$ by~\cite{PUZ20}
\beq
m \ddot \DE(t) = 2 d T - \b \int_0^t \de u M(t-u) \dot \DE(u) \ ,
\eeq
which in Laplace space, using $s^2 \wt \DE(s) = 2 d \wt Z(s)$, reads
\beq\label{eq:MvsZ}
m \wt Z(s) = T/s - \b \wt M(s) \wt Z(s)/s \qquad \Rightarrow \qquad \wt M(s) = \frac{T - m s \wt Z(s)}{\b \wt Z(s)} \ .
\eeq
Kinetic theory~\cite{DEC81,BMD85,LB89,DE04} gives the short-time expansion of $Z(t)$ in arbitrary dimension as
\beq
\frac{Z(t)}{Z(0)} = 1 - \frac{2|t|}{d\, \t_E} + A_d \left(\frac{t}{\t_E}\right)^2 + \OO(t^3) \ ,
\qquad
\t_E = \frac{\sqrt{\pi \b m}}{d\, 2^d\f} \frac{\ell}{g(\ell)} \ ,
\qquad
A_d = \frac{2}{d^2} {}_2F_1\left(-\frac12,-\frac12;\frac12(d+2);\frac14\right) \ ,
\eeq
where $Z(0)=T/m$. Moving to Laplace space and plugging this expansion in Eq.~\eqref{eq:MvsZ}
we get
\beq
\wt M(s) = \frac{2 T m}{d\, \t_E}\left[ 1 + \frac{2(1-A_d)}{d\, \t_E}\frac{1}{s} + \OO\left(\frac1{s^2}\right)\right] \ .
\eeq
This result shows that $M(t)$ is indeed the sum of a delta function and a regular function, which admits a short time expansion in integer powers of $t$.
Taking the $d\to\io$ limit with the rescaling $\wh m = (\ell^2/2 d^2) m$, $2^d \f = d\wh\f$, $\MM(t) =(\ell^2/2 d^2)M(t)$, and using $g(\ell)\to 1$ and $A_d = 1 + \a/d + \cdots$, we obtain
\beq
\wt\MM(s) = \wh\f \sqrt{\frac{2 T \wh m}\pi} \left[1 - \frac{2\a}{d \, \t}\frac{1}{s} 
+ \OO\left(\frac1{s^2}\right)\right] \ ,
\qquad
\t = d \, \t_E = \frac{\sqrt{2 \pi \b\wh m}}{\wh\f} \ .
\eeq
This result proves Eq.~\eqref{eq:Mshorttimeexp} when transformed back to the time domain. In particular, the coefficient of the delta peak coincides with the result obtained from DMFT, and the value of $\MM_{\rm reg}(t=0)$ is found to vanish proportionally to $1/d$ for $d\to\io$. Unfortunately, to our knowledge the next term of the short-time expansion has not been computed in finite $d$.

\subsection{Short-time expansion of the regular part of the memory function within DMFT}
\label{app:NewtonianHS2}

The regular part of the memory function $\MM_{\text{reg}} (t)$ for Newtonian hard spheres can be 
analytically computed for short times. Keeping only the leading short-time singularity $\MM(t) = 2 \z_0 
\d(t)$ as given in Eq.~\eqref{eq:Mdelta-HS}, the evolution of $h(t)$ reads
\beq\label{eq:Newt_HS_2}
\begin{split}
	\ddot h(t) + \z_0 \dot h(t) &= 1 + \X(t)  \ , \quad h(t)>0\\
	h(t=0) &= 0 \ , \quad \dot h(t=0) = g_0 \ , \\
	\langle \X(t) \X(u) \rangle  &= 2 \z_0 \d(t-u) \ .
\end{split}
\eeq
Because this white noise dominates at short times, we expect that in order to obtain the short-time 
behavior of $\MM_{\text{reg}}(t)$ we can neglect the regular part in the stochastic process, \ie we use the 
process in Eq.~\eqref{eq:Newt_HS_2}. Furthermore, at short times we can approximate the return probability by the {\it first} return probability, \ie neglect multiple returns.
The computation of the first return probability density  
$f(g_1,t|g_0)$ defined in Eq.~\eqref{eq:f-RTP} is a Wang-Uhlenbeck recurrence time problem.
Note that in the absence of the term $\zeta_0 \dot h(t)$, 
Eq.~\eqref{eq:Newt_HS_2} reduces to the random acceleration process in the presence of a linear drift~\cite{Bu07}, for which the first passage time distribution can be computed exactly~\cite{Bu08}. In presence of the linear drift, 
the problem
can be solved analytically for short times. In our units, we obtain~\cite{SS05} (see also~\cite{McK62,Bu08})
\beq\label{eq:f-IBM}
f(g_1,t|g_0) \sim \frac{\sqrt{3} \, |g_1|}{2\pi \z_0 t^2} \exp \left( \frac{g^2_0-g^2_1}{4} - \frac{g_0-g_1}{2\z_0} - \frac{g_1^2 + g_1 g_0 + g_0^2}{\z_0 t} \right) \erf \left( \sqrt{\frac{3 
|g_1| g_0}{\z_0 t}} \right) .
\eeq
The latter result can be plugged into Eq.~\eqref{eq:MregNewt3}, and the integral can be performed 
with the change of variables $u_i = |g_i|/\sqrt{\z_0 t}$ for $i=0,1$ and expanding around $t=0$. 
Setting now $\z_0 = \sqrt{2/\p} \wh\f$, one finds
\beq\label{eq:MregNewt_short}
\MM_{\text{reg}} (t) \sim \frac{\sqrt{6}}{\pi^{5/2}} \left[ \int^\io_0 \de u_0 \int_0^{\io} \de u_1 
u_0^2 \, u_1^2 \, e^{-(u_1^2-u_0 u_1 + u_0^2)} \, \erf \left( \sqrt{3 u_1 u_0} \right) \right] \, \wh\f^3 \, t
= C_0 \, \wh\f^3 \, t  \ ,
\eeq
with $C_0 \simeq 0.1578$. This result confirms that $B=0$
and $C = C_0 \, \wh\f^3$ when $d=\io$ in Eq.~\eqref{eq:Mshorttimeexp}. This short-time behavior has been plotted in 
Fig.~\ref{fig:HSN}a, showing a good agreement with the numerical solution of DMFT.

\subsection{Low-density regime of the DMFT equations}
\label{app:NewtonianHS3}

When $\wh\f \ll 1$, the regular part of the memory function is negligible, therefore $\MM(t) = 
2 \z_0 \, \d(t)$. Using Eq.~\eqref{eq:MSD} in dimensionless units,
with the above-mentioned memory kernel and $\wh\z=0$ one finds the solution
\beq\label{eq:MSD-HSN-lowphi}
\D(t) = \frac{1}{\z_0} \left[ t - \frac{1}{\z_0}  \left( 1 - e^{-\z_0 \, t} \right) \right] \ ,
\eeq
corresponding to the characteristic MSD of underdamped, dilute dynamics, with a ballistic regime 
for $t \ll \z_0^{-1}$ and a diffusive regime for $t \gg \z_0^{-1}$.
This allows one to compute analytically the viscosity and diffusion constant.
The two integrals in Eq.~\eqref{eq:SER-num} can be explicitly computed and give 
$\int_0^\io \de u \, \ddot \D (u)^2 = (2\z_0)^{-1} $ and $\int_0^\io \de u \, \MM (u) = \z_0 $.
Recalling that $ \z_0 = \sqrt{2/\p} \, \wh\f$, one finds
\beq\label{eq:SER-lowphi}
\b \wh D \wh\h_s = \wh\f \left( 1 + \frac{\p}{4 d \wh\f^2} \right) \ ,
\eeq
leading to the dilute-limit divergence $\b \wh D \wh\h_s \approx \p / (4 d \, \wh\f)$ when 
$\wh\f \to 0$ in any finite dimension. This divergence disappears when $d\to\io$.

\clearpage 
\bibliography{HSmerge.bib}

\end{document}